\documentclass[floats,floatfix,showpacs,amssymb,prd,twocolumn,superscriptaddress,nofootinbib,nolongbibliography,reprint]{revtex4-2}

\usepackage{amssymb,amsmath,verbatim,mathtools,needspace,enumitem,etoolbox,graphicx,physics,microtype,afterpage,bigints,gensymb,tabularx,xspace}
\usepackage{ragged2e}

\usepackage{multirow}
\usepackage[dvipsnames, usenames]{xcolor}
\definecolor{linkcolor}{rgb}{0.0,0.3,0.5}
\definecolor{dodgerblue}{HTML}{1E90FF}
\usepackage[unicode, colorlinks=true, linkcolor=linkcolor, citecolor=linkcolor, filecolor=linkcolor,urlcolor=linkcolor, pdfusetitle]{hyperref}
\usepackage[all]{hypcap}
\usepackage[T1]{fontenc}
\usepackage[utf8]{inputenc}
\usepackage{orcidlink}
\usepackage{subcaption}
\usepackage{ulem}
\usepackage{booktabs}
\usepackage[utf8]{inputenc}
\usepackage[T1]{fontenc}

\makeatletter
\newcommand*{\balancecolsandclearpage}{\close@column@grid \cleardoublepage \twocolumngrid}
\makeatother

\usepackage{xcolor}

\newif\ifshowcomments
\showcommentstrue 

\newcommand{\UPC}{\affiliation{Université Paris Cité, CNRS, Astroparticule et Cosmologie, F-75013 Paris, France}}

\begin{document}

\title{Parameter Estimation for Eccentric Supermassive Black Hole Binaries with Pulsar Timing Arrays}

\author{Sara Manzini}
\email{manzini@apc.in2p3.fr}

\author{Stanislav Babak}
\email{stas@apc.in2p3.fr}

\UPC

\date{\today}

\begin{abstract}

Pulsar timing array (PTA) experiments are searching for gravitational waves (GWs) in the nanohertz band. The primary GW sources targeted by PTAs include populations of inspiralling supermassive black hole binaries (SMBHBs) in the local Universe, some of which may emit detectable continuous gravitational-wave (CGW) signals. In this paper, we focus on the detection strategy for individual binaries on eccentric orbits. Using simulated datasets based on the EPTA DR2new configuration, we perform injection-recovery studies across the parameter space. We demonstrate that individual component masses can be measured when an eccentric SMBHB is detected at high GW frequencies. We further show a correlation between the CGW signal and the stochastic gravitational-wave background (SGWB) at low frequencies, which makes CGW identification challenging. Finally, the developed software is GPU-compatible, enabling efficient Bayesian inference. This work lays the groundwork for future applications to real PTA data.

\end{abstract}

\maketitle

\section{Introduction}

Gravitational wave (GW) astronomy entered the nanohertz regime with the recent evidence for an emerging GW signal reported by several Pulsar Timing Array (PTA) collaborations, including the European PTA (EPTA) + the Indian PTA InPTA \cite{EPTA:2023fyk}, NANOGrav \cite{NANOGrav:2023gor} and the Parkes PTA (PPTA) \cite{Reardon:2023gzh}, members of the International PTA (IPTA) consortium \cite{Verbiest:2016vem}. While the origin of this signal remains under active investigation, supermassive black hole binaries (SMBHBs) at sub-parsec separations are considered its most probable source. The SMBHBs on the broad orbits emit continuous gravitational waves (CGWs) and their incoherent superposition forms a stochastic GW signal at low (below a few tens of nanoHz) frequencies. Some particularly close and massive binaries could be individually resolved -- these CGW sources represent the next major scientific milestone for current and future PTAs. The prospects of their detection improve rapidly with the increase in the array size (the number of monitored pulsars) and the observational time span. 

The prospects of detecting CGWs from individual SMBHBs with current and future PTAs have been explored extensively through population-based forecasts \cite{Kelley:2018fur,Rosado:2015epa, Truant:2024aci, Truant:2025ybm}. These studies suggest that the next-generation of instruments such as the SKA will likely resolve individual sources above the background. Crucially, the detectability and parameter estimation accuracy of such sources depend sensitively on the assumptions about SMBHBs eccentricity and rely on accurate modelling of GW signals.

SMBHBs are formed following galaxy mergers, during which the black hole binary hardens through dynamical friction and interactions with stars and gas. When these environmental effects dominate over GW-driven circularisation, binaries can retain substantial eccentricity as they enter the PTA band.  Modelling GW emission from eccentric binaries was the main subject of our previous work reported in \cite{Manzini:2025gjx}. We demonstrated that eccentric binary waveforms introduce significant additional complexity compared to the circular case: (i) eccentric binaries evolve significantly faster than circular; (ii) gravitational radiation is distributed across multiple harmonics of the orbital frequencies (azimuthal and precession). Moreover, we argued that the use of the waveform for the circular binary (a very common assumption) will not be able to detect the eccentric signal even for mild eccentricity.

The main focus of this paper is to estimate the accuracy with which we can measure the parameters of eccentric binaries. Note that we do not consider here the detectability of eccentric CGWs (eCGWs); this will be the subject of a separate publication. We select binaries with sufficiently high, and therefore potentially detectable, signal-to-noise ratios (SNRs) from several simulated astrophysical populations \cite{Truant:2025ybm}. The first investigation concerns the conjecture presented in \cite{Manzini:2025gjx}: that we can obtain unbiased estimates of the masses of the individual SMBHs in an eccentric binary. This becomes possible thanks to the inclusion of periapse precession and post-leading-order relativistic terms in binary evolution. For the second investigation, we simulate data similar to EPTA DR2 \cite{EPTA:2023fyk} and examine (i) the accuracy of parameter estimation in different regimes (high and low frequency and high and low eccentricity) and (ii) the correlation between noise components and the eCGW signal.

eCGWs introduce additional parameters associated with eccentricity: the initial eccentricity, mass ratio, precession frequency, and initial precession phase. Including the pulsar term requires numerical integration of the ODEs describing the binary evolution. All these factors significantly increase the computational cost of Bayesian parameter inference for eCGWs. To keep the computations feasible, we use the \texttt{Discovery} package \cite{Vallisneri:2024xfk} to fit both the signal and noise parameters, as it natively supports GPU computations. The likelihood is written in \texttt{JAX}, and we use the JAX-compatible parallel-tempering ensemble Markov Chain Monte Carlo (MCMC) sampler \texttt{jexplore} \cite{jexplore}. Together, these tools allow us to exploit GPU hardware acceleration, reducing the computation time by approximately an order of magnitude while also lowering the carbon footprint compared to CPU-based computations.

The paper is organised as follows. In Section~\ref{sec:gwaveform} we give a brief overview of eCGW model. We describe the Bayesian framework adopted in the paper in Section~\ref{sec:data_analysis}. In Section~\ref{sec:validation}, we demonstrate the importance of pulsar term and post-Newtonian corrections in recovering parameters of eccentric SMBHBs. We investigate the correlation between noise components and eCGW in Section~\ref{sec:noiseCGW}. We summarize and discuss our findings in concluding Section~\ref{sec:discussion}.

\section{Eccentric binary model}\label{sec:gwaveform}

We refer the reader to \cite{Manzini:2025gjx} for a detailed description of the model; here we provide only a summary of its key aspects for completeness.

The timing residuals induced by eCGW for pulsar $\alpha$ are given as
\begin{equation}
\begin{split}
    r_\alpha(t) = &\left[F^+_\alpha\cos 2\psi 
                 + F^\times_\alpha\sin 2\psi\right]
                 \left[r_+(t) - r_+(t-\tau_\alpha)\right] \\
                -&\left[F^+_\alpha\sin 2\psi 
                 - F^\times_\alpha\cos 2\psi\right]
                 \left[r_\times(t) - r_\times(t-\tau_\alpha)\right],
\end{split}
\end{equation}
where $\psi$ is the GW polarisation angle and the antenna pattern functions for the $\alpha$-th pulsar are:
\begin{subequations}
\begin{align}
    F^+_\alpha &= \frac{(\hat{m}\cdot\hat{p}_\alpha)^2 
                - (\hat{n}\cdot\hat{p}_\alpha)^2}
                {2(1+\hat{\Omega}_{\rm GW}\cdot\hat{p}_\alpha)}, \\
    F^\times_\alpha &= \frac{(\hat{m}\cdot\hat{p}_\alpha)
                     (\hat{n}\cdot\hat{p}_\alpha)}
                     {1+\hat{\Omega}_{\rm GW}\cdot\hat{p}_\alpha}.
\end{align}
\end{subequations}
Here we introduced $\hat{p}_\alpha$ as the unit vector toward the pulsar, 
$\hat{\Omega}_{\rm GW}$ is the GW propagation direction, and
\begin{subequations}
\begin{align}
    \hat{m} &= (-\sin\phi_{\rm GW},\,\cos\phi_{\rm GW},\,0), \\
    \hat{n} &= (-\cos\theta_{\rm GW}\cos\phi_{\rm GW},\,
               -\cos\theta_{\rm GW}\sin\phi_{\rm GW},\,
                \sin\theta_{\rm GW}), \\
    \hat{\Omega}_{\rm GW} &= (-\sin\theta_{\rm GW}\cos\phi_{\rm GW},\,
                              -\sin\theta_{\rm GW}\sin\phi_{\rm GW},\,
                              -\cos\theta_{\rm GW}).
\end{align}
\end{subequations}
The two polarizations of the residuals $r_{+,\times}$ are obtained from the time integral of the GW strain polarizations $h_{+,\times}$:
\begin{subequations}
\begin{align}
    r_+(t) &= A(e, x)\,\Bigl\{(1+\cos^2\iota)\left[a(e,\xi)\cos 2\gamma + b(e,\xi)\sin 2\gamma\right] \notag\\
           &\phantom{{}= A(e, x)\,\Bigl\{} + \sin^2\iota\,c(e,\xi)\Bigr\}, \\[6pt]
    r_\times(t) &= 2A(e, x)\cos\iota\left[a(e,\xi)\sin 2\gamma - b(e,\xi)\cos 2\gamma\right],
\end{align}
\end{subequations}
with 
\begin{equation}
\begin{split}
    A(e, x) &= \frac{\nu M^2(1-e^2)^{3/2}}{D_L\sqrt{x}(1-e^2-3x)}\\
    a(e, \xi) &= \frac{(e+2\cos\xi)\sin\xi}{1+e\cos\xi}\\
    b(e, \xi) &= \frac{\cos 2\xi+e\cos\xi}{1+e\cos\xi}\\
    c(e, \xi) &= \frac{e\sin\xi}{1+e\cos\xi}
\end{split}
\end{equation}
where $\iota$ is the inclination of the orbital angular
momentum to the line of sight, $D_L$ is the luminosity distance and $\nu = m_1 m_2/M^2$ is the symmetric mass ratio. The phases $\xi(t)$ (true anomaly) and $\gamma(t)$ (periapsis advance) describe conservative orbital dynamics, while $e(t)$ (eccentricity) and $x(t)$ evolve adiabatically under radiation reaction. The variable $x$ is defined in terms of the azimuthal orbital frequency as $x = (M\omega_\phi)^{2/3}$.

The two contributions are evaluated at Earth time ($t$) and at the retarded pulsar time ($t - \tau_\alpha$), where the light-travel time from the pulsar to Earth is given by $\tau_\alpha = L_\alpha\left(1 + \hat{\Omega}_{\rm GW}\cdot\hat{p}_\alpha\right)$. These will be referred to as the \emph{Earth term} and \emph{pulsar term}, respectively.

In our previous paper~\cite{Manzini:2025gjx}, we explicitly presented the equations governing the binary evolution. In particular, we emphasize the importance of post-leading-order terms when evolving the orbit over the light-travel time between the Earth and pulsar terms. The dynamics are described by four time-dependent variables: $\xi(t)$, $\gamma(t)$, $x(t)$, and $e(t)$. In total, the eCGW is described by 11 parameters: two initial phases, namely the true anomaly and the precession phase; the initial eccentricity, defined at the Earth time $t=0$; the initial orbital frequency, associated with the azimuthal phase, $\omega_{\phi} = 2\pi F_{\rm{orb}}$; the total mass $M$; the symmetric mass ratio $\nu$; the source sky position $(\theta_{\rm{GW}}, \phi_{\rm{GW}})$; the orbital inclination $\iota$; the luminosity distance $D_L$; and the polarization phase $\psi$. Including the pulsar term in the residuals introduces $2N_{\rm{psr}}$ extra parameters: one additional pulsar-term phase (true anomaly), and one pulsar distance for each pulsar in the array. The distances to the pulsars are generally poorly known, and we allow them to vary according to Gaussian priors with standard deviations equal to 20\% of their nominal values, unless a different accuracy is explicitly specified.

\section{Bayesian Framework}\label{sec:data_analysis}

The main aim of this paper is to demonstrate our ability to estimate the parameters of eccentric SMBHBs using PTA observations. We adopt a Bayesian framework in which the parameters are treated as random variables and their posterior distributions are inferred from simulated datasets. In the following subsections, we describe the PTA data model and the implementation of the Bayesian framework.

\subsection{PTA data model}
We begin by introducing the PTA data model.  
The TOAs of a single pulsar (indexed as $\alpha$) can be written as \cite{vanHaasteren:2012hj}:
\begin{equation}
    {t}^{arr}_{\alpha} = {t}^{TM}_{\alpha} + {\delta t}_{\alpha}
\end{equation}
The deterministic term includes the timing model (TM) which accounts for inaccuracies in the pulsar's sky location, proper-motion, spin evolution, the dispersion measure (DM) and its first two derivatives \cite{EPTA:2021fqa}. We also account for the orbital motion of the pulsar if it is in a binary system.
An initial fit for the TM parameters is performed using the least-squares ($\chi^2$) method applied to the ToAs and provides the maximum likelihood parameters $\beta_0$. Those estimates do not take into account a possible GW signal or the time-correlated noise components described below. However, we assume that the initial fit is close enough to the true values so that the residual errors  $\epsilon = \beta - \beta_{0}$ can be described by a linear model with the design matrix, $M\epsilon$. 

The design matrix $M$ is an $(n \times m)$ matrix, where $n$ is the number of observed TOAs and $m$ is the number of TM parameters ($\beta$).

Several noise components are usually present in the PTA observations. The white noise (WN) corresponds to the instrumental (and, in general, measurement) noise. The individual (pulsar-dependent) red noise (RN) is associated with stochastic variation in the spin of the pulsars. We also have the so-called "chromatic" red noise -- dispersion measure variations (DMv) and scattering variations (Sv) — which depend on the radio-frequency of observation and correspond to time-varying interactions of electromagnetic waves with interstellar plasma. The SGWB is present in the observations as a correlated common red-noise process, with correlations depending on the angular separation between pulsars on the sky and described by the Hellings--Downs curve~\cite{Hellings:1983fr}. However, the correlated component of the noise is significantly weaker than 
the diagonal components of the cross-pulsar noise spectrum.

In this work, we focus on the eCGW signal and therefore simplify the simulated data model. First, we mimic the SGWB as a common 
\emph{uncorrelated} red-noise process (CRN), effectively neglecting the correlated component of the SGWB in order to reduce computational cost (this allows us to factorize the likelihood over pulsars). The main conclusions of this paper are not affected by this approximation.
Second, we neglect individual red-noise (RN) components and chromatic noise because (i) this saves substantial computational time and 
(ii) our primary goal is to study the correlation between the eCGW signal and the noise, for which CRN is sufficient.
All in all, the data model used in this paper is described as 
\begin{equation}
    \delta{t} = M {\epsilon}  + 
    s_{\text{eCGW}} + n_{\text{WN}} + n_{\text{CRN}} .
\end{equation}
The first two components are deterministic and correspond to the linearised TM and eCGW signal. 

The white noise component is described by the power spectral density (PSD)
$S_{\text{WN}} = 2 \sigma^2\Delta t$, where $\sigma = \sigma_{\text{TOA}}$ 
corresponds to TOA timing uncertainty and $\Delta t$ is the typical observation cadence. 
The corresponding covariance matrix in the time domain is diagonal: 
$N_{ij} = \delta_{ij}\sigma_i^2$.
The red noise is assumed to be diagonal in the frequency domain and is described by the 
power-law PSD:
\begin{equation}\label{eq:RN_PSD}
    S(f) = \frac{A_{\mathrm{crn}}^2}{12\pi^2}\left(\frac{f}{f_{\text{yr}}}\right)^{-\gamma_{\mathrm{crn}}}f_{\text{yr}}^{-3},
\end{equation}
 where $f_{\text{yr}} = yr^{-1}$.  Later in the paper, we fit the spectral content of the CRN in the frequency domain by applying the free-spectrum model \cite{Lentati:2016ygu}. The measurements $\rho_i = \rho(f_i)$ used in the free spectrum are related to the PSD by $S(f_i, \rho_i) = \rho_i^2 T$. We choose frequencies $f_i = i/T$, corresponding to the Fourier bins, and restrict the analysis to the first 30 components, $i\le 30$.
The correponding covariance matrix in the time domain is obtained by applying the cosine transform (real Fourier Transform) of $S(f)$ and approximated as \cite{vanHaasteren:2012hj}: 
\begin{equation}\label{eq:C_RN}
\begin{split}
    C(\tau_{ij}) &= \left(\frac{A}{2\sqrt{3}\pi}\right)^2 \frac{1yr^{3-\gamma}}{f_L^{\gamma -1}} \Bigg\{\Gamma (1-\gamma)\sin\left(\frac{\pi \gamma}{2}\right)(f_L \tau_{ij})^{\gamma-1} \\&- \sum_{n=0}^{\infty}(-1)^n\frac{(f_l \tau_{ij})^{2n}}{(2n)!(2n +1 - \gamma)}\Bigg\}
\end{split}
\end{equation}
where $\tau_{ij} = 2\pi |t_i - t_j|$, $f_{\text{low}}$ is the cut-off frequency  and we truncate the sum at $n=2$.
Red noise dominates over white noise at low frequencies, implying that long-timescale correlations (large $\tau_{ij}$) are important. 

Given two different regimes of the dominant noise, we also consider low and high-frequency eCGW separately. We stress, however, that eccentric binaries generate multiple harmonics spanning a broad frequency range; therefore, the classification into “low” and “high” frequency refers to the initial azimuthal orbital frequency of the system.

\subsection{Bayesian inference}

The likelihood assumes that the residuals of each pulsar are described as Gaussian noise after subtracting the TM and eCGW signal:

\begin{eqnarray}
\ln \mathcal{L}_{\alpha} &=& -\frac{1}{2}\,
(\delta t_{\alpha} - s_{\alpha} - M_{\alpha}\epsilon_{\alpha})^{\top}
\mathbf{\Sigma}^{-1}_{\alpha}
(\delta t_{\alpha} - s_{\alpha} - M_{\alpha}\epsilon_{\alpha}) 
\nonumber \\
&-& n_{\alpha}\ln \pi - \ln \det{|\mathbf{\Sigma}_{\alpha}|},
\end{eqnarray}
where $n_{\alpha}$ is the number of observations, the noise covariance matrix is given by $\Sigma_{ij} = N_{ij} + C_{ij}$, and $s_{\alpha} = s_{\alpha}(\Theta)$ describes the eCGW signal. Since we neglect the correlations between pulsars, the total log-likelihood is given by
\begin{equation}
\ln\mathcal{L} = \sum_{\alpha=1}^{N_{\mathrm{psr}}} \ln\mathcal{L}_{\alpha}.
\end{equation}

\subsubsection{Frequency Domain Likelihood}
We can compute the likelihood more efficiently by approximating the CRN as ~\cite{Lentati:2012xb, vanHaasteren:2014qva}:

\begin{equation}\label{eq:C_WN_CRN}
\mathbf{\Sigma}_{\alpha} = \mathbf{N}_{\alpha} + \mathbf{F}_{\alpha} \mathbf{\Phi} \mathbf{F}_{\alpha}^T,
\end{equation}
where $F_{\alpha} = \{\sin{\omega_i t_{\alpha}},\,\, \cos{\omega_i t_{\alpha}}  \}$ are the partial Fourier basis functions and 
$\Phi_{ij} = S(f_i)\delta_{ij}/T$, assuming that the frequencies are uncorrelated.\footnote{It was shown in~\cite{Crisostomi:2025vue} that the frequency components are correlated and that this correlation should be taken into account in real data analyses. Here, we neglect these correlations since our focus is on the eCGW signal.}

The inversion of this covariance matrix can be performed efficiently using the Woodbury matrix identity:
\begin{equation}    
    \mathbf{\Sigma}_{\alpha}^{-1} = \mathbf{N}_{\alpha}^{-1} - \mathbf{N}_{\alpha}^{-1}\mathbf{F}_{\alpha}
    (\mathbf{\Phi}_{\alpha}^{-1}+\mathbf{F}_{\alpha}^{T}\mathbf{N}_{\alpha}^{-1}\mathbf{F}_{\alpha})^{-1}
    \mathbf{F}_{\alpha}^T\mathbf{N}_{\alpha}^{-1},
\end{equation}
and its determinant is given by
\begin{equation}
    \ln|\mathbf{\Sigma}_{\alpha}|= \ln|\mathbf{N}_{\alpha}| + \ln|\mathbf{\Phi}| + 
    \ln|\mathbf{\Phi}^{-1} + \mathbf{F}_{\alpha}^{T}\mathbf{N}_{\alpha}^{-1}\mathbf{F}_{\alpha}|.
\end{equation}

This form of the likelihood is implemented in \texttt{JAX} in the \texttt{Discovery} package~\cite{Discovery_github} and is used in the most computationally demanding parts of our analysis. 

\subsubsection{Time-Domain Likelihood}

The likelihood can be evaluated in the time domain; however, this approach involves two computationally expensive steps. The first is the computation of $C_{ij}(\tau)$ from Eq.~\eqref{eq:C_RN} and the inversion of the full covariance matrix $\mathbf{\Sigma}_{\alpha}^{-1}$, which constitutes the main computational bottleneck. The second is the evaluation of the eCGW template over a large number of samples. Since PTA datasets typically have dimensionality $N_{\mathrm{TOAs}} \sim 10^3 - 10^4$, one possible strategy for reducing the computational cost is to downsample the dataset, thereby making the time-domain likelihood more tractable. This could be a valid path to investigate the low-frequency content of PTA data.

We test this possibility by retaining one residual out of every $D$ consecutive TOAs. This reduces the dataset size by a factor of $D$ and lowers the effective Nyquist frequency from $f_{\mathrm{Nyq}}$ to $f_{\mathrm{Nyq}}/D$. A direct consequence of this operation is aliasing: signal components above the new Nyquist frequency, $f_{\mathrm{Nyq}}/D$, are folded back into the retained frequency band, potentially distorting the reconstructed signal. To suppress this effect, we apply a low-pass Butterworth filter to both the residuals and the covariance matrix before downsampling, attenuating power above $f_{\mathrm{Nyq}}/D$. This procedure is valid only when all dominant harmonics of the eCGW signal lie well below $f_{\mathrm{Nyq}}/D$, ensuring that downsampling does not degrade the total SNR.

As a proof of principle, we analysed a simulated dataset consisting of a low-frequency eCGW with $e_0 = 0.5$, $\log_{10}F_{\mathrm{orb}} = 12,\mathrm{nHz}$, and $\log_{10}(M/M_{\odot})=9.56$, together with WN characterised by $\sigma_{\mathrm{TOAs}} = 2.5,\mu\mathrm{s}$ and $\mathrm{EFAC}=1$, as well as CRN with $\log_{10}A_{\mathrm{CRN}} = -14.5$ and $\gamma_{\mathrm{CRN}} = 4.33$. The first inference was performed using the original evenly sampled dataset with a cadence of $\Delta t_{\rm orig}=15$ days and a duration of 10 years. The data were then downsampled to a cadence of $\Delta t_{\rm ds}=90$ days. In both cases, the SNR of the injected signal was 5.7, and the red-noise parameters were fixed to their injected values. The resulting posteriors for the eCGW parameters were statistically consistent, while the run time was reduced by a factor of 2 due to the smaller number of samples entering the computation of the signal-induced timing residuals and the likelihood inner products. A further speed-up, scaling approximately as $(\Delta t_{\rm ds}/\Delta t_{\rm orig})^3$, is expected from the matrix inversion required in a joint analysis of the CRN/GWB and eCGW signals.

Real PTA datasets are more complex. The TOAs are unevenly sampled in time, and the noise budget includes chromatic components, such as dispersion-measure variations and scattering noise. As a result, combining TOAs from different frequency channels and applying a uniform downsampling scheme is non-trivial, and requires a more careful treatment. 

\subsubsection{Posteriors}
We infer the posterior distribution of the data-model parameters (eCGW and noise) within a Bayesian framework:
\begin{equation}
    p(\Theta, \lambda | \delta t, \mathcal{M}_a) =
    \frac{\mathcal{L}(\delta t|\Theta,\lambda, \mathcal{M}_a)\,
    \pi(\Theta,\lambda |\mathcal{M}_a)}
    {p(\delta t|\mathcal{M}_a)}.
    \label{eq:BayesPE}
\end{equation}
Here, $\mathcal{M}_a$ denotes the assumed data model, where we separate the parameters describing the eCGW signal ($\Theta$) from those describing the noise ($\lambda$). The prior distribution of these parameters is given by $\pi(\Theta,\lambda |\mathcal{M}_a)$, and $p(\delta t|\mathcal{M}_a)$ is evidence for the model $\mathcal{M}_a$. 

The evidence,
\begin{equation}
    p(\delta t|\mathcal{M}_a) = \int d\Theta\, d\lambda \,
    \mathcal{L}(\delta t|\Theta,\lambda, \mathcal{M}_a)
    \pi(\Theta,\lambda |\mathcal{M}_a),
\end{equation}
is used to quantify the relative support for different models through the odds ratio:
\begin{equation}
    \mathcal{O}_{a,b} =
    \frac{p(\mathcal{M}_a|\delta t)}
    {p(\mathcal{M}_b|\delta t)}
    =
    \frac{p(\delta t|\mathcal{M}_a)}
    {p(\delta t|\mathcal{M}_b)}
    \frac{\pi(\mathcal{M}_a)}
    {\pi(\mathcal{M}_b)}
    =
    \text{BF}_{a,b}
    \frac{\pi(\mathcal{M}_a)}
    {\pi(\mathcal{M}_b)},
\end{equation}
where $\text{BF}_{a,b}$ is the Bayes factor. For equal prior probabilities assigned to the models, the Bayes factor quantifies which model is better supported by the data. Note, however, that it does not determine whether a model is intrinsically good; rather, it provides a relative ranking between competing models.

Inference of model parameters is typically performed numerically using Markov chain Monte Carlo (MCMC) methods~\cite{Ivezic:2014}. Computing the evidence is challenging, particularly in high-dimensional parameter spaces such as the one considered here. The most common approach is nested sampling~\cite{Skilling:2004pqw}. However, the Bayes factor can also be evaluated without explicitly computing the evidence using reversible-jump MCMC (RJMCMC)~\cite{Littenberg:2023xpl, Karnesis:2023ras} or the product-space approach~\cite{Hee:2015eba, Bartolucci:2006}. Alternatively, one can use normalizing flows to construct a neural density estimator based on posterior samples and estimate the evidence from the base distribution and Jacobians of the transformations~\cite{Vallisneri:2024xfk}.

We use a particular MCMC implementation: \texttt{Jexplore} (\url{https://jexplore-773e5b.pages.in2p3.fr/README.html}), an ensemble sampler with parallel tempering (using several walkers per temperature). The package is written in \texttt{JAX} and can be run on both CPU and GPU architectures. We vectorize the likelihood evaluation over all walkers and temperatures, which significantly improves the efficiency of the sampler. Additionally, \texttt{Jexplore} can evaluate the Bayes factor using the product-space technique.

\section{Results}

This section is split into two parts. In the first part, we investigate the estimation of companion masses in the eccentric binary
and the effect of high-order PN terms. We also emphasise the importance of pulsar terms in eCGW. The second part is dedicated to (almost) 
realistic PTA settings where we investigate inference (and partially detectability) of eCGWs in the presence of a CRN. In particular, we focus on the correlation between common red noise (mimicking the SGWB) and the GW signal from an individual eccentric MBHB. 

\subsection{Estimating MBH masses in the eccentric binaries}\label{sec:validation}
\begin{table}[b]
\centering
\begin{tabular}{@{}lll@{}}
\toprule
\textbf{Data} & \textbf{Model} & \textbf{Figure} \\
\midrule
\multirow{3}{*}{\shortstack[l]{WN + eCGW\\ ($F_{\mathrm{orb}}=15$ nHz, $e_0=0.5$)}}
  & CGW                               & \ref{fig:corner_1vs2PN}, \ref{fig:corner_5sigmapd}, \ref{fig:corner_ET_highF} (teal) \\
  & CGW (leading order) & \ref{fig:corner_1vs2PN} (pink) \\
  & CGW (ET only)                     & \ref{fig:corner_ET_highF} (yellow) \\
\cmidrule(l){1-3}
\multirow{2}{*}{\shortstack[l]{WN + eCGW\\ ($F_{\mathrm{orb}}=5$ nHz, $e_0=0.01$)}}
  & CGW           & \ref{fig:corner_circ_EPT} (teal) \\
  & CGW (ET only) & \ref{fig:corner_circ_EPT} (yellow) \\
\bottomrule
\end{tabular}
\caption{Simulated datasets and models used in the inference.}
\label{tab:data_model1}
\end{table}
Here we employ a simple data model, which is sufficient to serve our purposes. The considered PTA consists of six pulsars used in~\cite{EPTA:2021crs, Speri:2022tua}, selected to maximize the SNR while keeping the computational cost low. These are J0613-0200, J1012+5307, J1600-3053, J1713+0747, J1744-1134, and J1909-3744 all placed at $1\mathrm{kpc}$ distance. The TOAs are evenly spaced with a two-week cadence over an observation time span of $T = 20\,\mathrm{yr}$. We assume that all pulsar observations contain only white noise with the same rms timing uncertainty, $\sigma_{\mathrm{TOA}} = 2.5\,\mu\mathrm{s}$. This is an idealized dataset aimed at demonstrating the importance of pulsar term and higher-order PN terms in estimating the individual black hole masses in the binary. In addition, we use these data to investigate the impact of uncertainties in pulsar distances in eCGW paramaters estimation.

Restricting the noise to a white component allows us to study the impact of the full harmonic structure of the eCGW signal on the timing residuals. Using six pulsars demonstrates our ability to measure the individual masses even with a small array while keeping the computational cost low.

We produce several datasets each containing an eCGW signal with parameters chosen from the simulated astrophysical population 
described in  \cite{Truant:2025ybm}. The datasets are summarised in Table~\ref{tab:data_model1}.
We inject eCGW containing both Earth and pulsar terms at the highest PN order described in \cite{Manzini:2025gjx}. The SNR of the source against WN is 10.52. We inject a binary with chirp mass 
$\log_{10}(\mathcal{M}/M_\odot) = 9.2$, 
initial eccentricity $e_0 = 0.5$, 
orbital frequency $F_{\rm orb} = 14.85 \rm{nHz}$,
and luminosity distance $\log_{10}(d/{\rm Mpc}) = 1.2$.
For the inference we are using several models: (i) eCGW model the same as injected, (ii) eCGW at leading order approximation, neglecting all high (post-leading order) PN corrections, (iii) eCGW with the Earth term only. In all models we assume that WN level is known. In addition, we also study the impact of pulsar distance uncertainty on the parameter estimation, considering two Gaussian priors 
with $\sigma_{\mathrm{pdist}} = 20\%$ (our feducial assumption) and 
$\sigma_{\mathrm{pdist}} = 5\%$.

\begin{figure}[htbp]
    \centering
    \begin{subfigure}[b]{0.48\textwidth}
        \includegraphics[width=\textwidth]{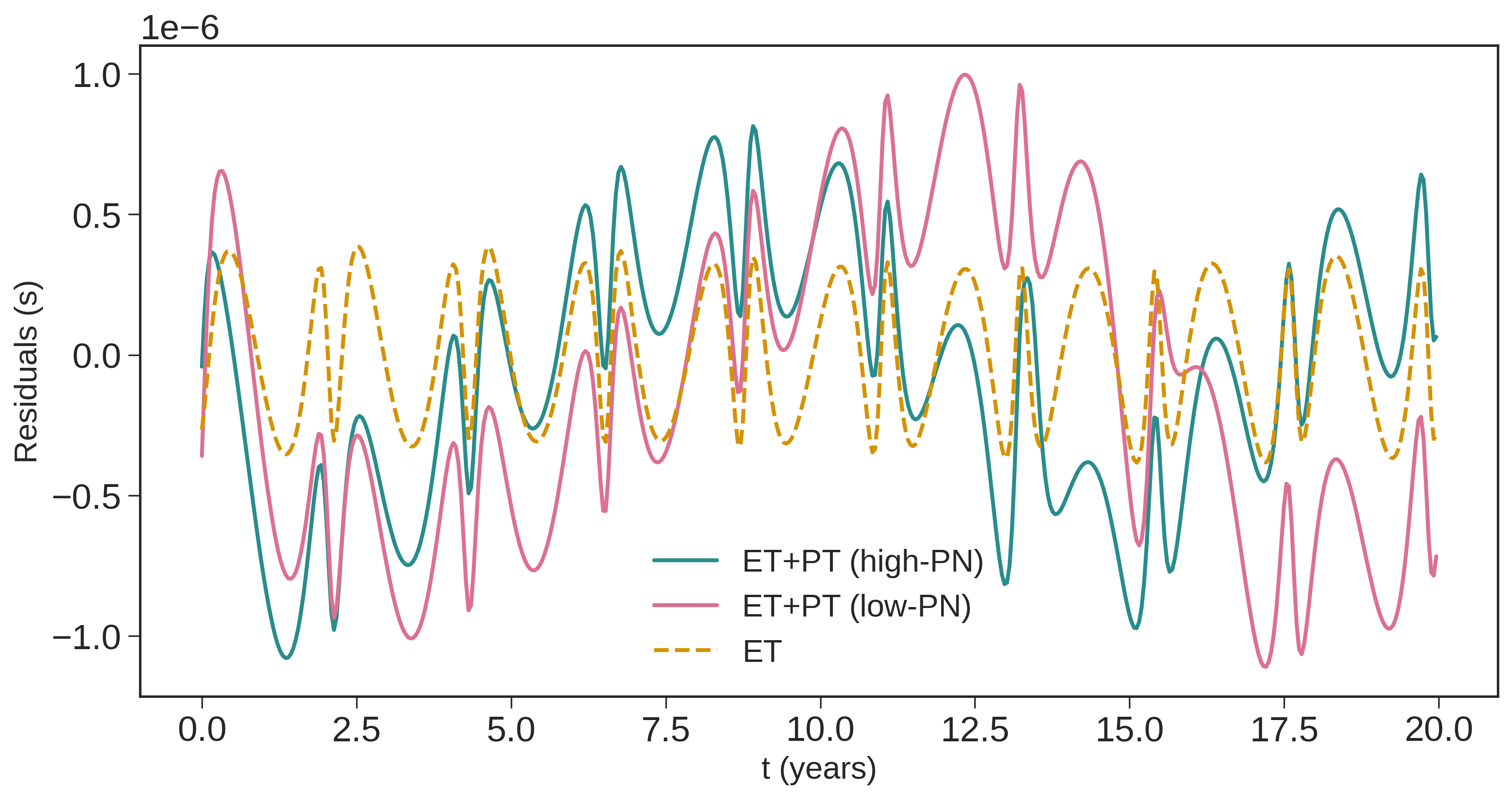}
        \caption{Time domain residuals including ET and PT for a binary with initial eccentricity $e_0 = 0.5$ and initial orbital frequency $F_{\mathrm{orb}} = 15 \mathrm{nHz}$, considering ET only (yellow), post leading order corrections in dynamics (teal) and leading order dynamics (pink).}
        \label{fig:res_e06F15}
    \end{subfigure}
    \hfill
    \begin{subfigure}[b]{0.48\textwidth}
        \includegraphics[width=\textwidth]{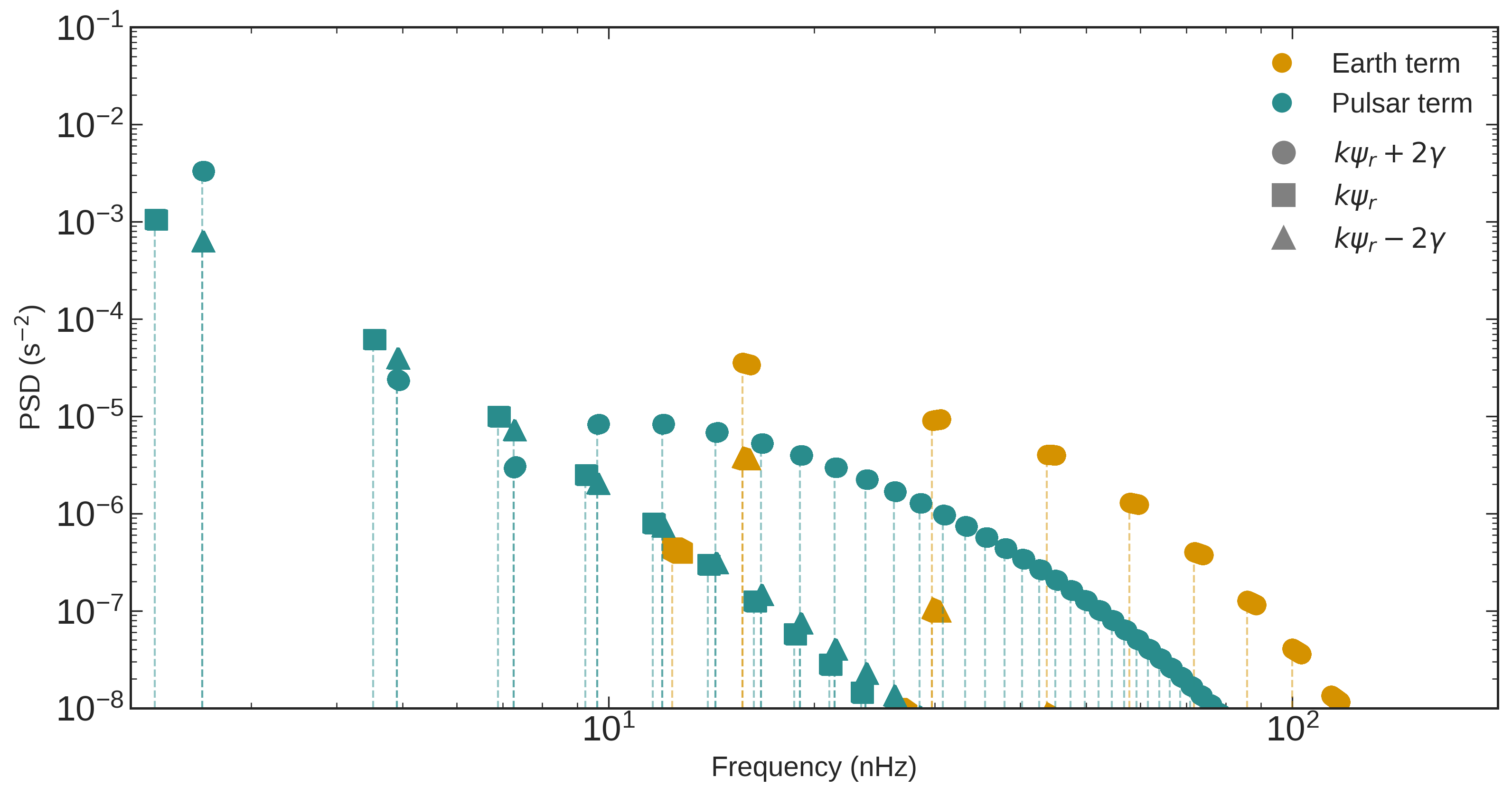}
        \caption{Frequency domain power spectral density for one pulsar including ET (yellow) and PT (teal) for the same binary as in Figure ~\ref{fig:res_e06F15}, considering post-leading-order corrections in dynamics. Different markers represent different sets of harmonics: $k\psi_r + 2\gamma$ are circles, $k\psi_r$ squares, and $k\psi_r - 2\gamma$ triangles.}
        \label{fig:spectrum_e06F15}
    \end{subfigure}
        \caption{Residuals and spectral analysis for eccentric high-frequency and mild eccentricity binary.}
    \label{fig:overall}
    \vspace{0.5em}
\end{figure}

The relatively high starting frequency $F_{\text{orb}}=15nHz$ justifies using white noise only in the simulated data, as all Earth terms will lie in the white noise dominated frequency range. However, the pulsar terms appear between 1 and 10 nHz.  The spectrum of the simulated eCGW signal is shown in the Fig.~\ref{fig:spectrum_e06F15}. We also observe a small evolution of the orbital frequency over the observation span assumed to be $20 \mathrm{yrs}$.

\begin{figure}[htbp]
    \centering
    \begin{subfigure}[b]{0.48\textwidth}
        \includegraphics[width=\textwidth]{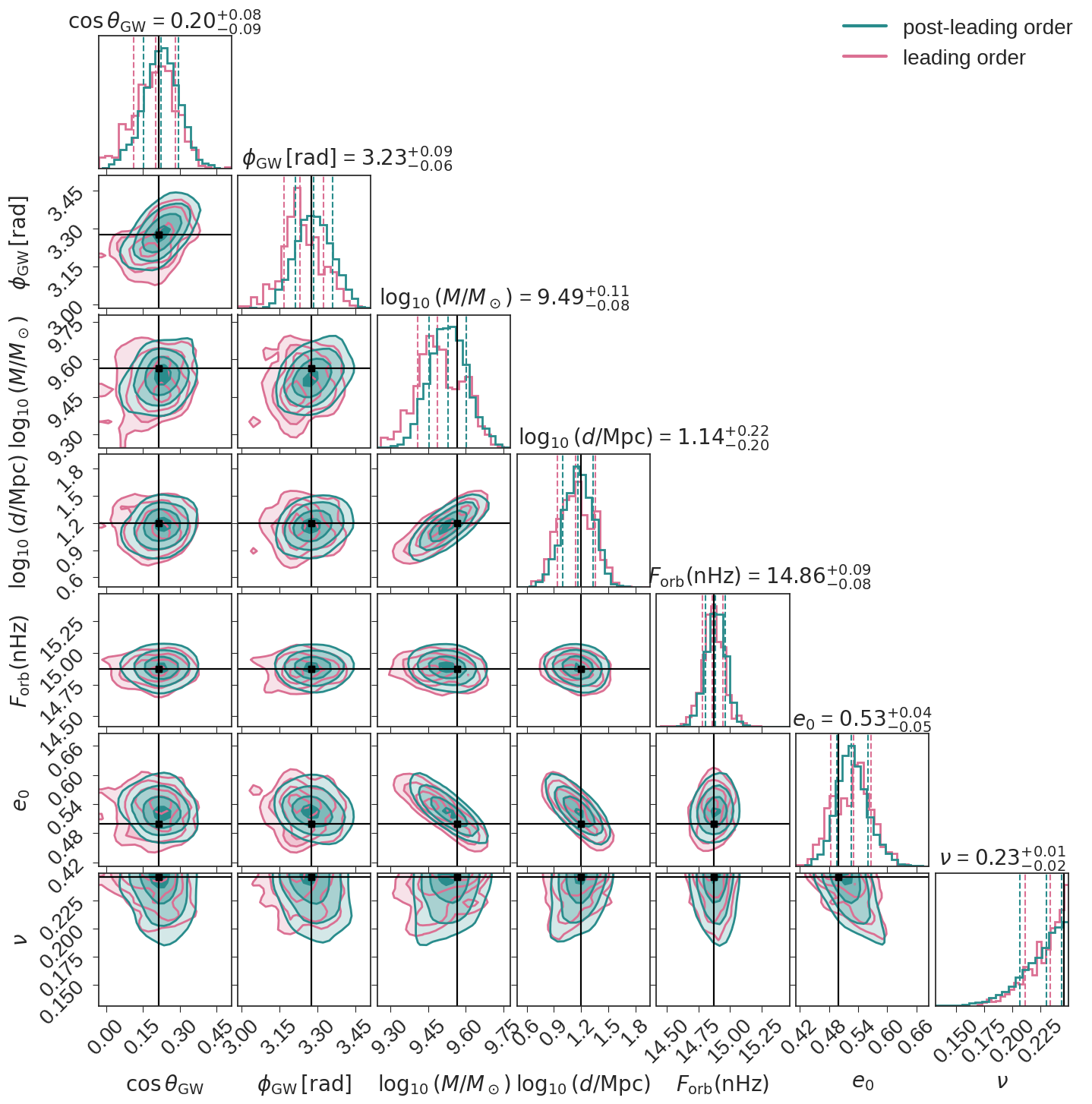}
        \caption{eCGW parameters posteriors}
        \label{fig:corner_1vs2PN}
    \end{subfigure}
    \hfill
    \begin{subfigure}[b]{0.48\textwidth}
        \includegraphics[width=\textwidth]{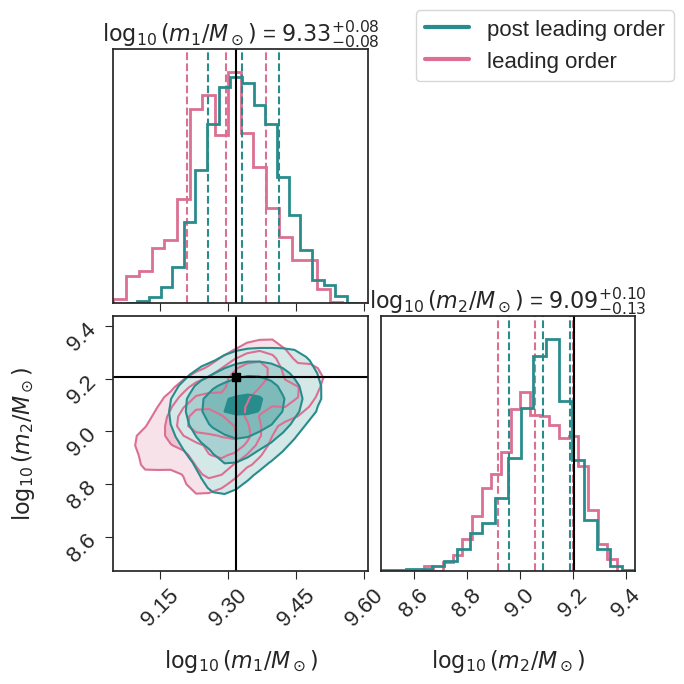}
        \caption{Posterior distribution for $m_1, m_2$ obtained by combining the posteriors of $\log_{10}M$ and $\nu$ from Fig.~\ref{fig:corner_1vs2PN}.}
        \label{fig:corner_m1m2}
    \end{subfigure}
    \caption{Posterior distribution for eCGW parameters for the binary at $F_{\rm orb} = 14.85 \rm{nHz}$ and eccentricity $e_0 = 0.5$. The posterior in teal is obtained by fitting with the full ET + PT waveform with dynamics at post-leading order, while the pink posterior is obtained with leading order model.}
    \label{fig:full_corner1vs2pn}
    \vspace{0.5em}
\end{figure}

\begin{figure}[h!] 
    \includegraphics[width=0.5\textwidth]{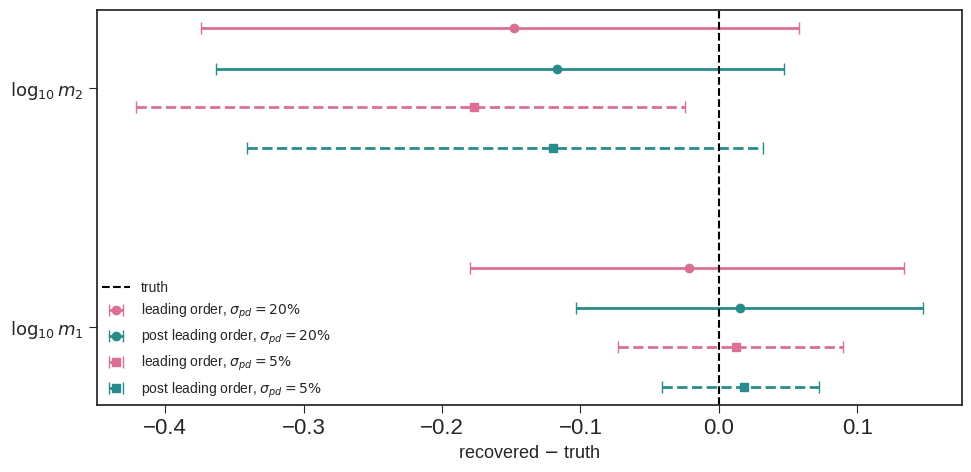}
    \caption{Discrepancy (median and $90\%$ credible interval) between recovered and true $m_1, m_2$ for leading-order waveform (pink) and post-leading order (teal) with $\sigma_{pd} = 20\%$ (circles and solid lines) and  $\sigma_{pd} = 5\%$ (squares and dashed lines).}
    \label{fig:m1_m2_difference}
\end{figure}

One of our goals is to demonstrate the importance of high-order PN terms in computing the pulsar terms, since the binary is expected to evolve significantly over $\tau_{\alpha}$. The difference between the residuals induced by an eCGW signal with and without high-order PN terms is visually apparent in Fig.~\ref{fig:res_e06F15}. However, this difference can be partially compensated by biases in the inferred parameters, as shown in the posterior distributions in Fig.~\ref{fig:corner_1vs2PN} and Fig.~\ref{fig:corner_m1m2}.  In this test, we placed all pulsars at a distance of $1\,  k\mathrm{pc}$ and assumed that this distance is known with a relative uncertainty of $20\%$.

The posteriors for the full (correct) eCGW model are given in teal in Fig.~\ref{fig:corner_1vs2PN}. We see informative posteriors for all parameters, including the symmetric mass ratio ($\nu$). This confirms that we can constrain the individual masses of an SMBHB (see Fig.~\ref{fig:corner_m1m2}). On the other hand, neglecting the PN terms in the binary evolution over $\tau_{\alpha}$ leads to (i) a drop in the SNR to 10.14, (ii) broader posteriors (in pink), and (iii) larger biases in the recovered parameters.

The biases partially compensate for the mismatch in the timing residuals, which is permitted by the large uncertainty in the pulsar distance. As a test, we repeated the analysis after reducing $\sigma_{\mathrm{pdist}}$ to a 5\% relative error. The corresponding results are shown as dashed lines in Fig.~\ref{fig:m1_m2_difference}. We clearly see that the leading-order model starts to fail in fitting the data, and the injected value of $m_2$ lies outside the $90\%$ credible interval.

\begin{figure}[h!] 
    \includegraphics[width=0.5\textwidth]{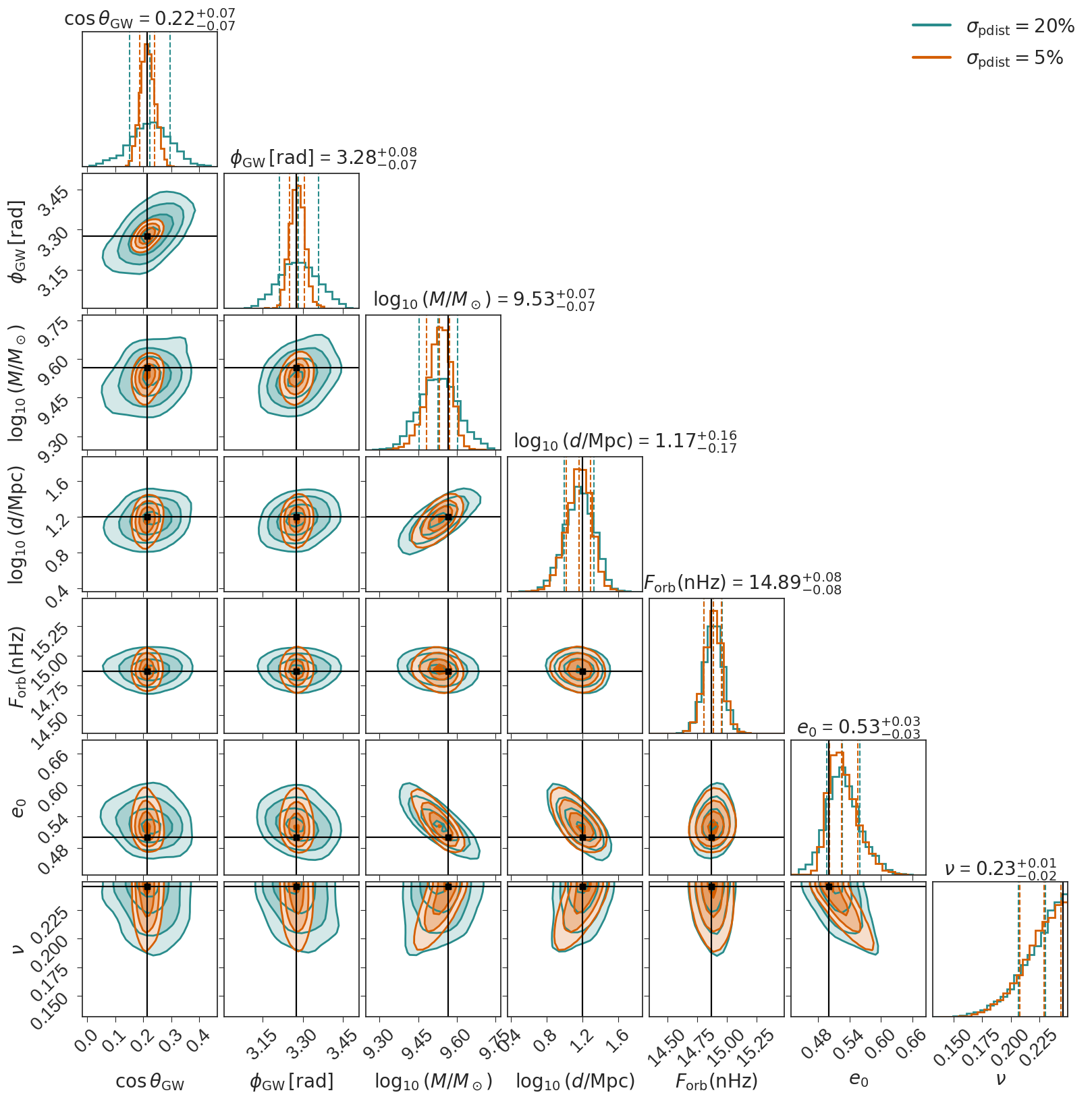}
    \caption{Posterior distribution for the model using a post-leading-order eCGW waveform and simulated residuals from six pulsars. The teal contours show the posteriors recovered using a pulsar-distance uncertainty of $\sigma_{pd} = 20\%$, while the orange contours correspond to $\sigma_{pd} = 5\%$.}
    \label{fig:corner_5sigmapd}
\end{figure}

Next, we explore the effect of the uncertainty in the pulsar distance on the parameter estimation of the eccentric binary.
Since the pulsar distances are inferred as part of Bayesian fitting, we observe correlations across pulsars, which are unphysical and can also affect the correlation between the other eCGW parameters, most notably those constrained by their evolution over $\tau_{\alpha}$. We reduce the uncertainty in the pulsar distances to $5\%$ while keeping the eCGW model the same as that used for the simulation.

The resulting posteriors are presented in Fig.~\ref{fig:corner_5sigmapd}. The teal posterior corresponds to our reference, and the orange contours represent the posteriors obtained with reduced uncertainty in the pulsar distances.
As expected, the biggest impact is on the sky localisation of the GW source, while the improvement in the estimation of the other parameters is negligible; still, we observe a slight change in the correlation between total mass and symmetric mass ratio.

Finally, we investigate the impact of the pulsar term: can we drop the pulsar term from the model?

\begin{figure}[t]
\centering

\begin{subfigure}{0.48\textwidth}
    \includegraphics[width=\textwidth]{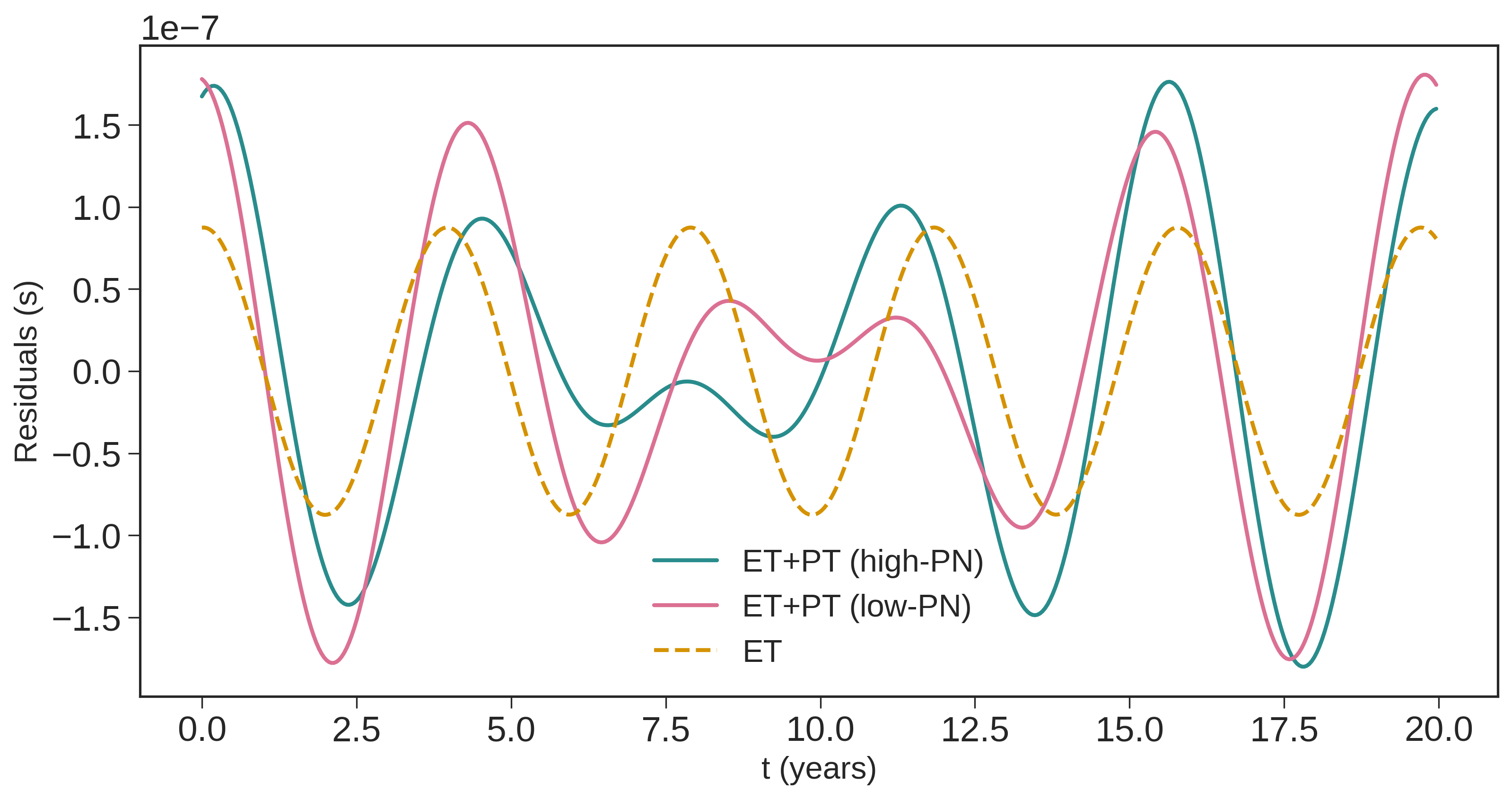}
    \caption{Time-domain residuals for a binary with initial eccentricity $e_0 = 0.003$ and initial orbital frequency $F_{\mathrm{orb}} = 4 \mathrm{nHz}$. The full waveform with post-leading-order corrections in the dynamics is given by teal line. The ET-only contribution is shown as a yellow dashed line, and 
    the leading-order dynamics waveform is in pink.}
    \label{fig:res_e003F4}
\end{subfigure}

\vspace{1.5em}  

\begin{subfigure}{0.48\textwidth}
    \includegraphics[width=\textwidth]{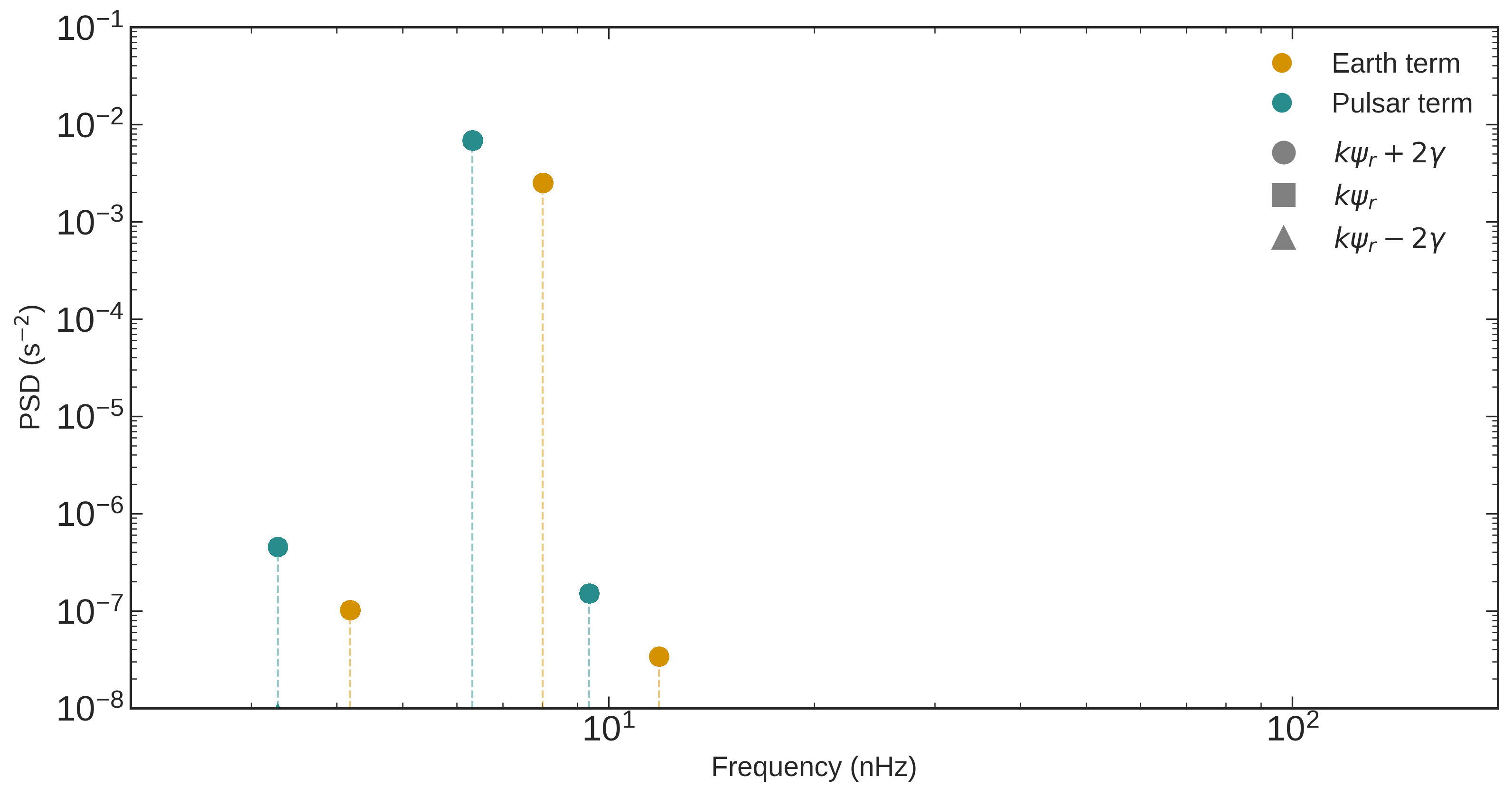}
    \caption{Frequency-domain power spectral density for one pulsar, including ET (yellow) and PT (teal), for the same binary as in Figure~\ref{fig:res_e003F4}, using post-leading-order corrections in the dynamics. Different markers represent different sets of harmonics: $k\psi_r + 2\gamma$ are circles, $k\psi_r$ squares, and $k\psi_r - 2\gamma$ triangles.}
    \label{fig:spectrum_e003F4}
\end{subfigure}

\caption{Residuals and spectral analysis for a low-frequency, low-eccentricity binary.}
\label{fig:overall_lowflowe}

\end{figure}

\begin{figure}[h!]
\centering

\begin{subfigure}[b]{0.48\textwidth}
    \includegraphics[width=\textwidth]{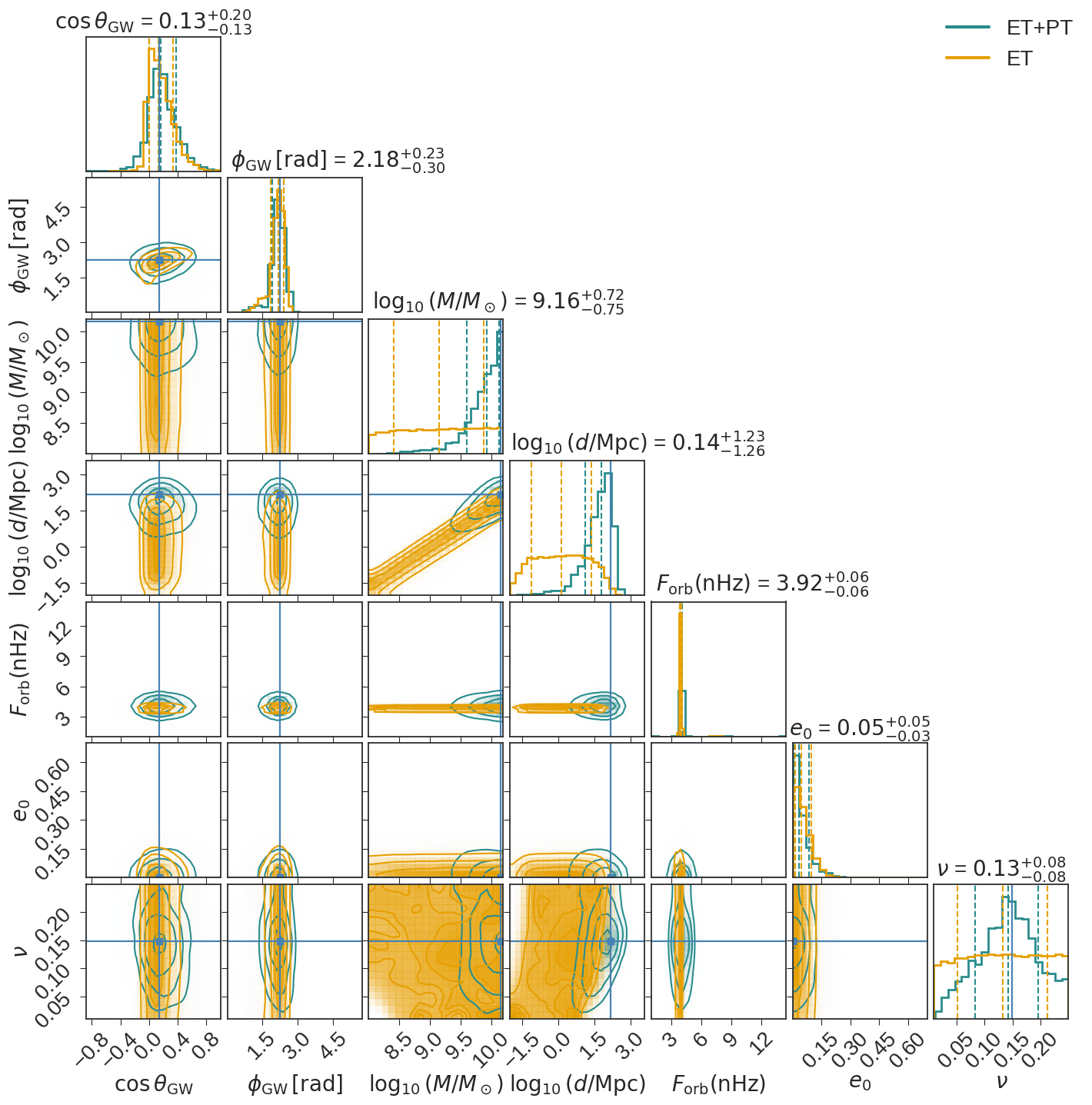} 
    \caption{Posterior distribution obtained for a dataset with an injected source at initial eccentricity $e_0 = 0.003$ and initial orbital frequency $F_{\mathrm{orb}} = 4 \mathrm{nHz}$.}
    \label{fig:corner_circ_EPT}
\end{subfigure}

\vfill
\begin{subfigure}[b]{0.48\textwidth}
    \includegraphics[width=\textwidth]{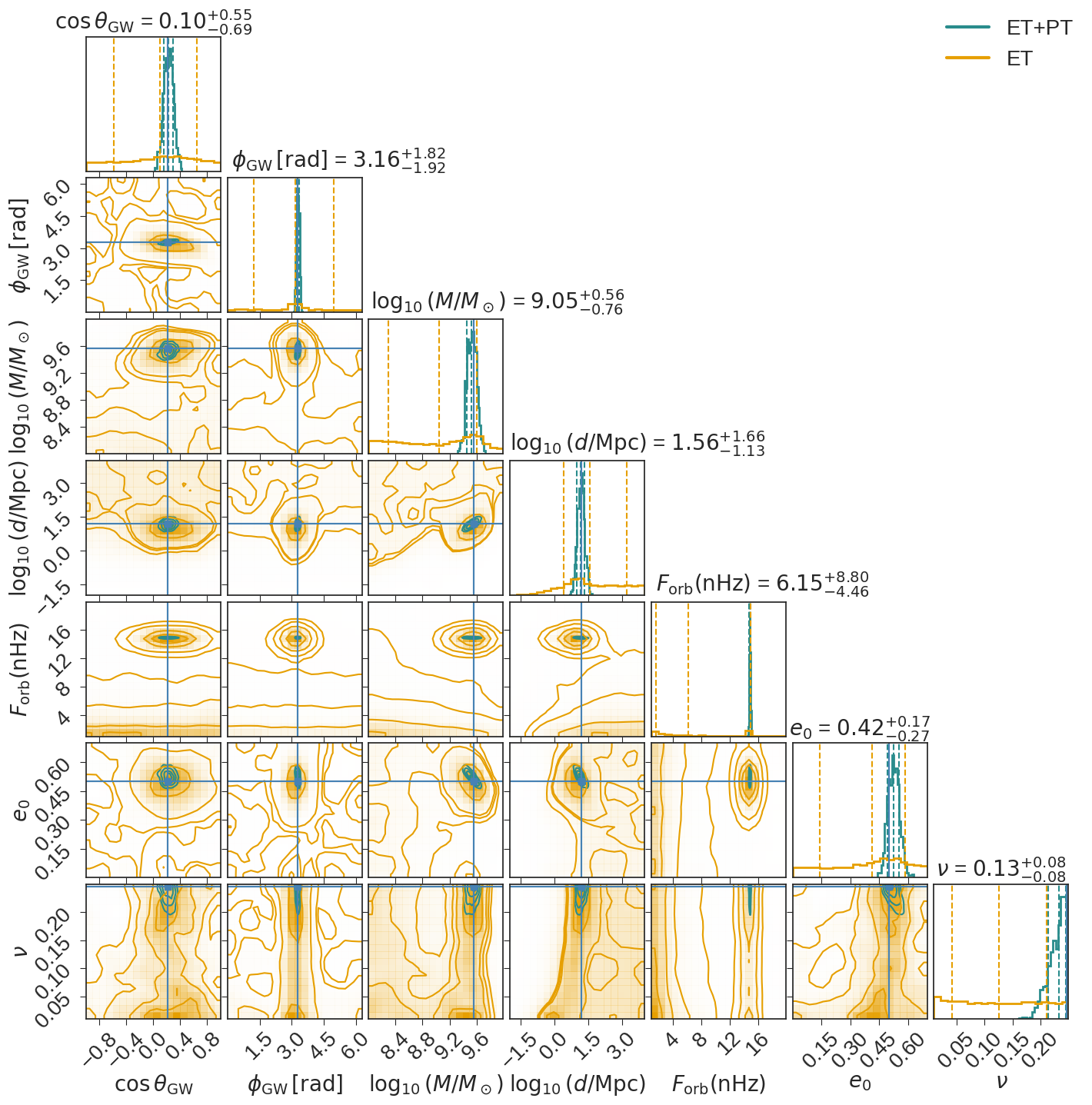} 
    \caption{Posterior distribution obtained for a dataset with an injected source at initial eccentricity $e_0 = 0.5$ and initial orbital frequency $F_{\mathrm{orb}} = 15 \mathrm{nHz}$.}
    \label{fig:corner_ET_highF}
\end{subfigure}
\caption{Comparison of posteriors obtained with a full model (both pulsar and Earth terms) (teal) and with the model with Earth term only (yellow). The simulated data contained the full eCGW signals including high PN corrections to the dynamics.}
\label{fig:corner_comparison_ET}

\end{figure}

First, we consider an almost circular binary ($e=0.003$) with initial orbital frequency $F_{\mathrm{orb}} = 4 \mathrm{nHz}$ and $\mathrm{SNR} = 8$. The residuals and the spectrum of harmonics for this source are presented in Fig.~\ref{fig:overall_lowflowe}.
This binary does not evolve significantly over $\tau_{\alpha}$, for example,  for pulsar J1744-1134, the eccentricity at the pulsar term is $e_0^p=0.004$, and the orbital frequency is $F_{\mathrm{orb}}^p = 3.16 \mathrm{nHz}$. As a result, neglecting the pulsar terms does not prevent us from detecting the source. The inferred posteriors with (teal) and without (yellow) the pulsar term in the model are given in Fig.~\ref{fig:corner_circ_EPT}. The full signal model gives well-constrained posteriors, while the Earth-term-only model leads to (i) unconstrained posteriors for the mass and mass ratio (which is expected since we lack frequency evolution), (ii) biases, most notably in the frequency and the initial phase, and (iii) broader posteriors for all parameters.

We observe a different picture for eCGWs at high frequency. The ET-only model is too simplistic to accurately reproduce the full harmonic structure of the eccentric signal, and consequently, all the eCGW parameters remain unconstrained. This is demonstrated by the yellow posteriors in Fig.~\ref{fig:corner_ET_highF}, which show substantial degradation compared to our fiducial posterior obtained using the ET+PT model (teal).

Dropping the pulsar term leads to a decrease in the overall optimal SNR of the sources similar for both the eccentric and circular binaries, it is reduced to
$\mathrm{SNR}_{\mathrm{ET}} \sim 5$. One the other hand the matched filtering SNR (inner product between the signal and the maximum likelihood estimator) decreases from 8 to 7.6 for the circular binary and from 10.52 to 5.7 for the eccentric SMBHB. This has direct observational consequences: dropping the pulsar term in the eccentric model might fail to detect the signal altogether.

The drop in SNR in yellow posteriors in Fig.~\ref{fig:corner_ET_highF} for eccentric binaries is a direct consequence of the multi-harmonic structure of eCGW. In contrast, for low-frequency and almost circular binaries, the groups of harmonics associated with the Earth and pulsar terms lie in the same frequency bins, since the binary evolution over $\tau_{\alpha}$ is still negligible, and the ET-only model can still capture the overall signal at the expense of biases.

\subsection{Correlation of eCGW with the CRN}\label{sec:noiseCGW}

\begin{table}[h!]
\centering
\begin{tabular}{@{}llc@{}}
\toprule
\textbf{Data} & \textbf{Model} & \textbf{Figure} \\
\midrule
\cmidrule(l){1-3}
\multirow{4}{*}{\shortstack[l]{CRN + CGW\\ ($F_{\mathrm{orb}}=3$ nHz, $e_0=0.75$)}}
  & CRN + CGW              & \ref{fig:cgw_corner_full}, \ref{fig:cgw_crn_corner} (cyan)\\
  & CRN (power law)        & \ref{fig:cgw_crn_corner} (purple)\\
  & CRN (freespectrum)     & \ref{fig:cgw_crn_spectrum} (cyan violins)\\
\cmidrule(l){1-3}
\multirow{3}{*}{\shortstack[l]{CRN + CGW\\ ($F_{\mathrm{orb}}=15$ nHz, $e_0=0.5$)}}
  & CRN + CGW    & \ref{fig:corner_CRN_CGW_highF}, \ref{fig:corner_crn_highF} (cyan)\\
  & CRN (power law)      & \ref{fig:corner_crn_highF} (purple)\\
  & CRN (freespectrum)   & \ref{fig:spectrum_highF} (cyan violins)\\
\bottomrule
\end{tabular}
\caption{Summary of different data sets and modeling approaches for CRN, and CGW signals. All the runs include WN both in the data set and the model, with parameters fixed to per backend values contained in the noise dictionaries.}
\label{tab:data_model2}
\end{table}

The main focus of this subsection is to understand and assess the correlation between noise and the eCGW signal.
To answer this question, we need to use more realistic simulated data. The data simulated in this subsection closely
resemble the EPTA DR2new dataset \cite{Manzini:2025gjx}. We use the actual epochs of observations and measurement errors for the TOAs of 25 pulsars, which implies non-uniformly sampled data. We add a CRN with $\log_{10}(A_{\mathrm{crn}}) = -14.5$ and $\gamma_{\mathrm{crn}} = 4.33$. We also inject an eCGW with a low orbital frequency $F_{\mathrm{orb}} = 3.15\,\rm{nHz}$ and high eccentricity $e_0 = 0.74$ to study the correlation between the CRN and the eCGW signal. The simulated eCGW signal has $\mathrm{SNR}_{\mathrm{eff}} = 8.89$. We use a highly eccentric binary to have GW power spread across many harmonics
in the frequency domain and potentially amplify the covariance between the noise and the signal. The summary of the simulated data sets and the models used in the inference is given in Table~\ref{tab:data_model2}.

\begin{figure}[t] 
    \includegraphics[width=0.5\textwidth]{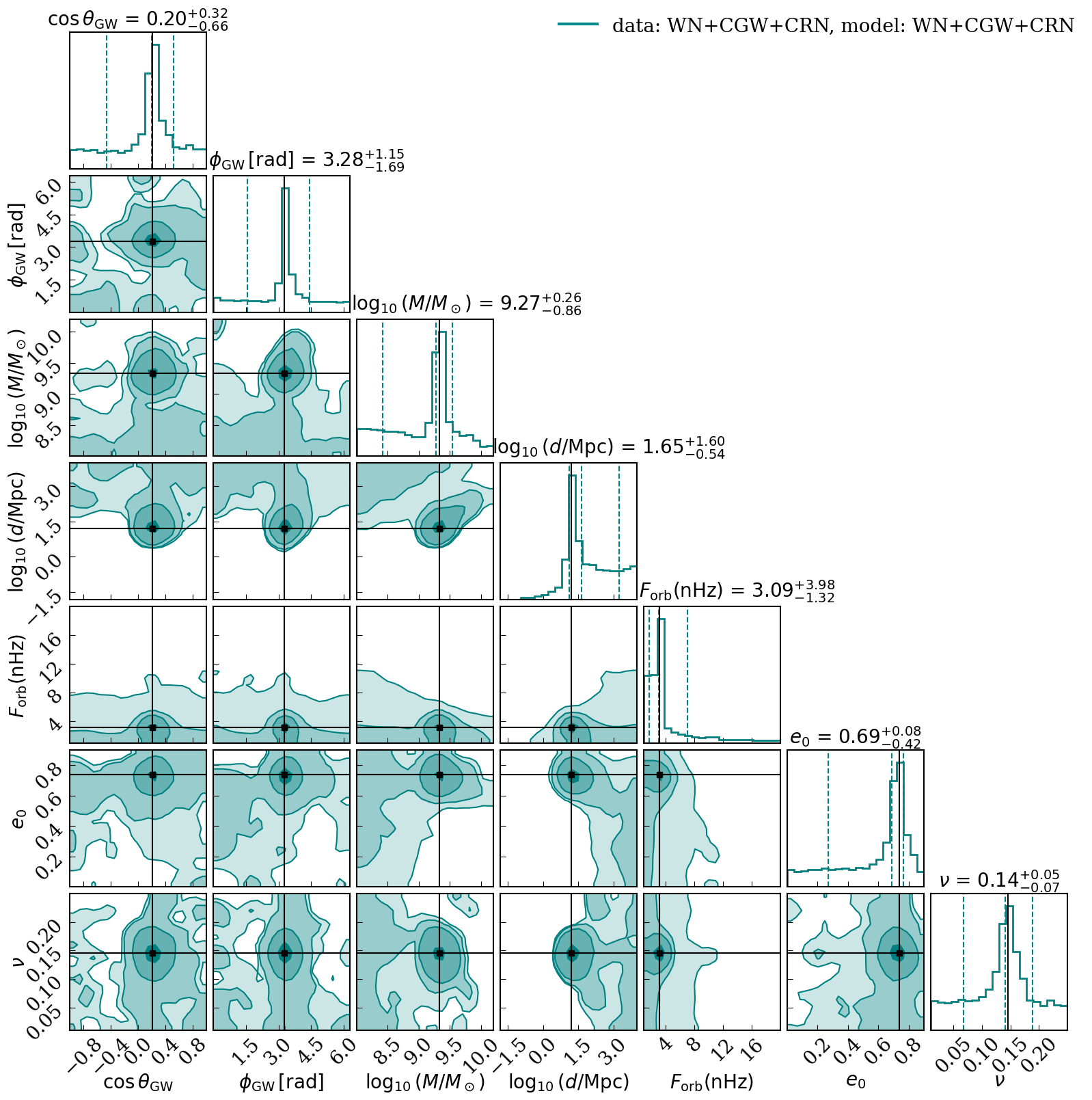} 
    \caption{Posterior distribution for CGW parameters using model containing WN+CGW+CRN on the WN+CGW+CRN dataset.}
    \label{fig:cgw_corner_full}
\end{figure}

\begin{figure}[t] 
    \includegraphics[width=0.5\textwidth]{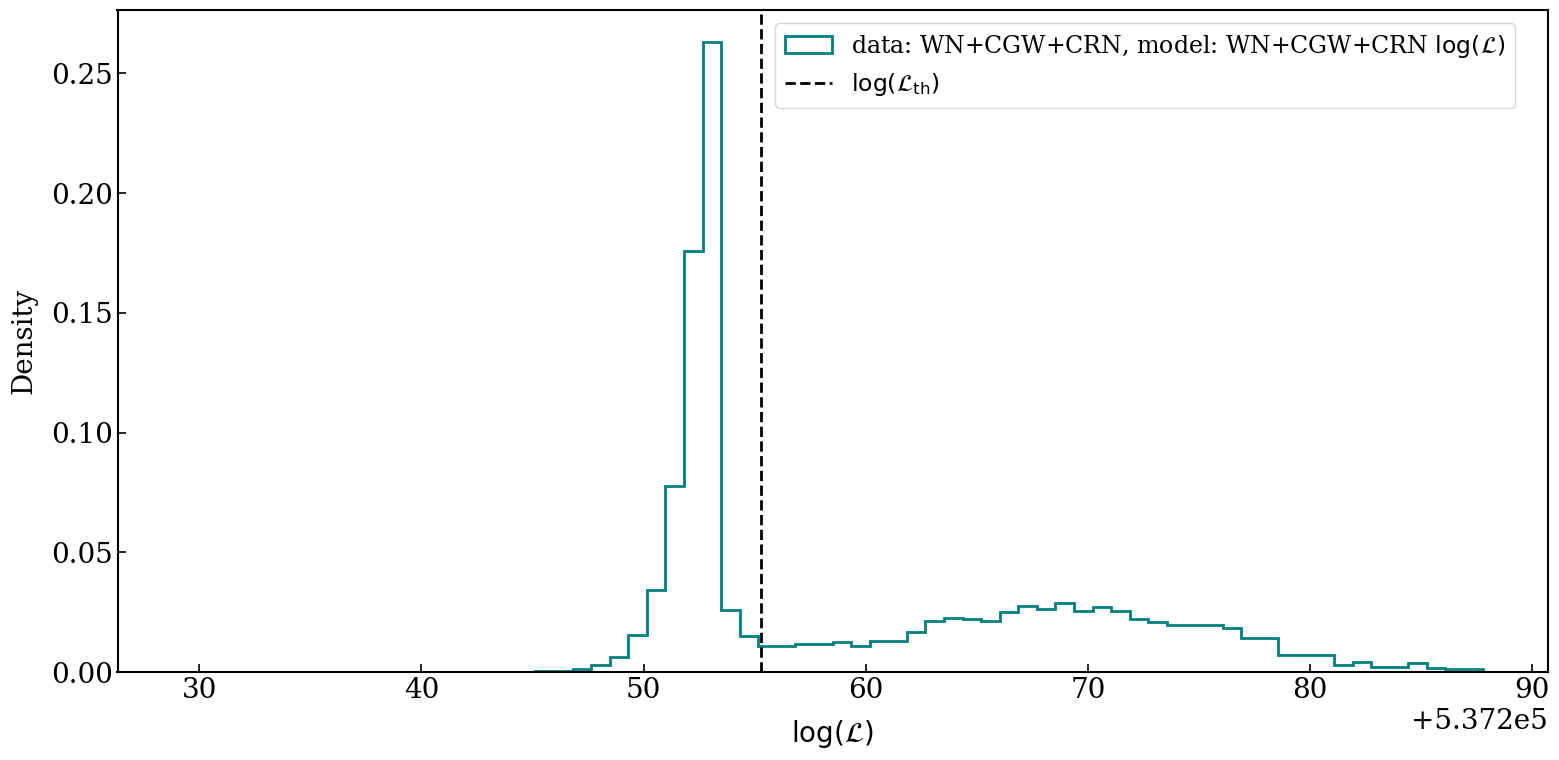} 
    \caption{Log-likelihood distribution for parameter estimation on data: WN+CRN+CGW, using model: WN+CRN+CGW (teal). The black dashed line is the threshold for splitting the posteriors in eCGW constrained and absorbed by the CRN. }
    \label{fig:cut_likelihood}
\end{figure}

\begin{figure}[t] 
    \includegraphics[width=0.5\textwidth]{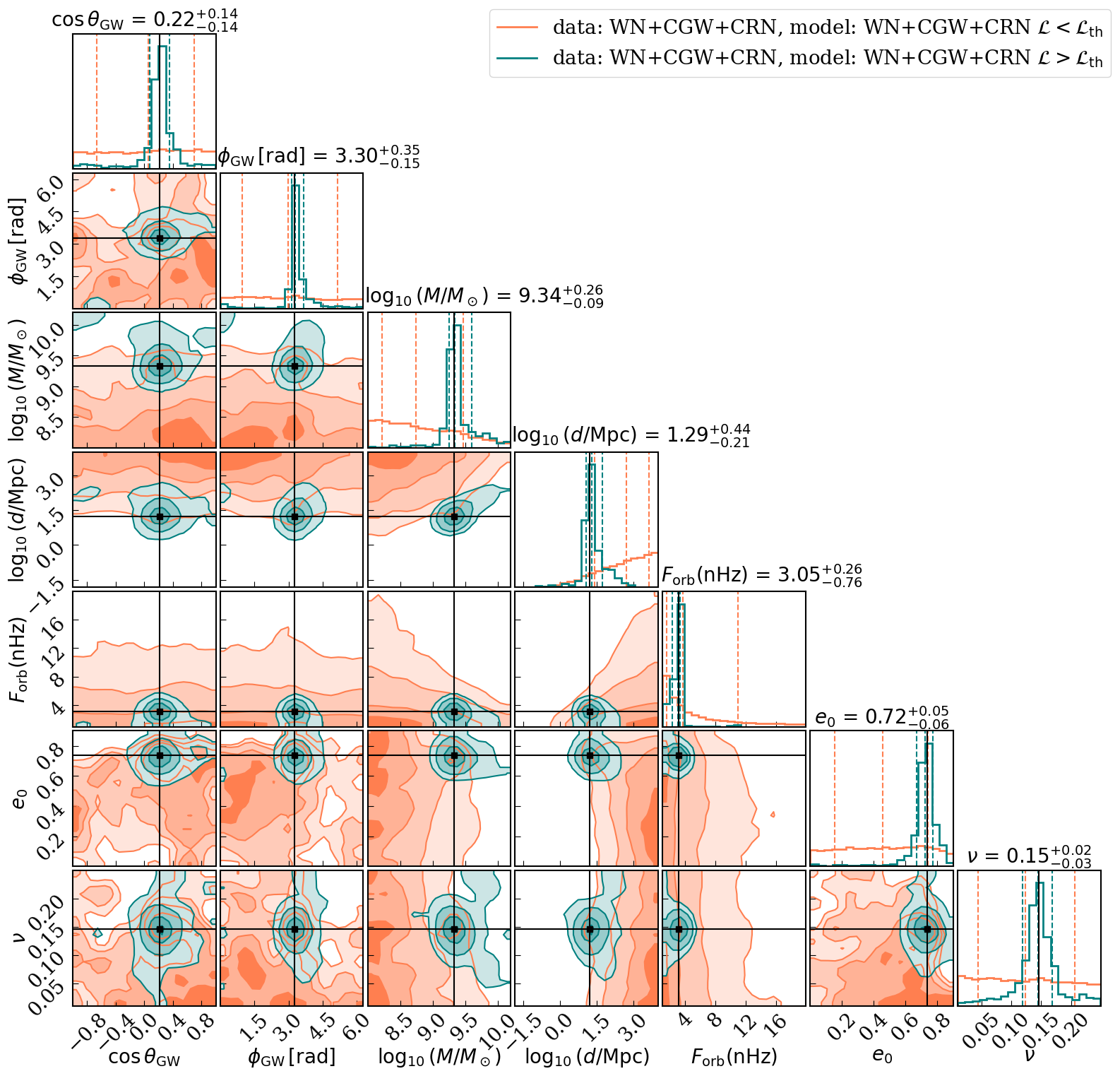} 
    \caption{Posterior distribution for CGW parameters using model containing WN+CGW+CRN on the WN+CGW+CRN dataset. Teal (orange) posterior are obtained selecting points with $\mathcal{L}>\mathcal{L}_{\mathrm{th}}$ ($\mathcal{L}<\mathcal{L}_{\mathrm{th}}$) }
    \label{fig:cgw_corner_th}
\end{figure}

\begin{figure}[b]
        \includegraphics[width=0.45\textwidth]{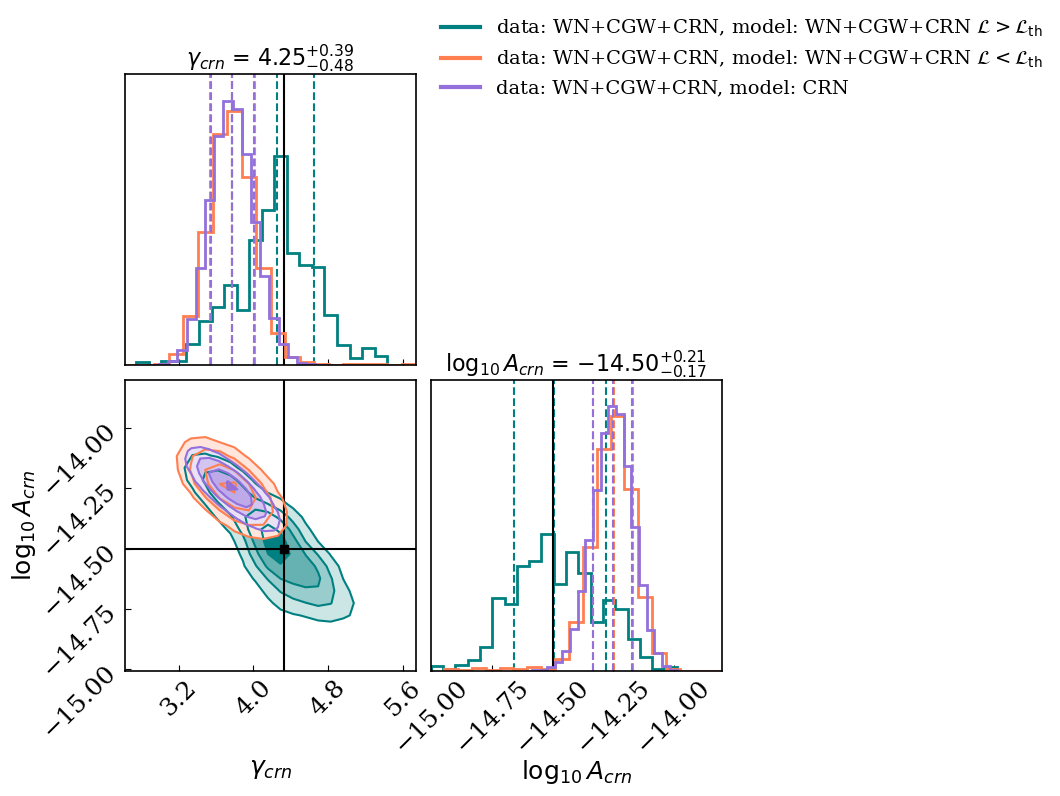}
        \caption{CRN parameters posterior distribution for different simulations. The posterior distribution in teal (orange) is obtained by fitting dataset with WN+CRN+CGW with model WN+CRN+CGW and selecting only points with $\mathcal{L}>\mathcal{L}_{\mathrm{th}}$ ($\mathcal{L}<\mathcal{L}_{\mathrm{th}}$). The purple posterior is obtaine fitting the same dataset with model WN+CRN.}
        \label{fig:cgw_crn_corner}
\end{figure}

We infer the parameters of the eccentric SMBHB together with the CRN parameters; both the simulated data and the model consist of CRN+WN+eCGW. The resulting posterior
is given in Fig.\ref{fig:cgw_corner_full}, and the corresponding distribution of the log-likelihood is presented in Fig.~\ref{fig:cut_likelihood}. 
The likelihood exhibits a bimodal structure. We split the posterior samples into low- and high-likelihood modes with $ \mathcal{L}_\mathrm{th} = 537255.3$ as the separator and replot the posterior using teal colour for the high-likelihood mode and orange for the low-likelihood mode in Fig.~\ref{fig:cgw_corner_th}. The CRN posterior is shown in Fig.~\ref{fig:cgw_crn_corner}.
We observe that the teal posterior corresponds to the actual data model and correctly recovers the parameters of the eCGW and CRN, while the orange posterior points correspond to a model with CRN  only, returning a non-informative posterior for the eCGW (upper bound).   Moreover, the orange posterior points correspond to a CRN that is flatter in slope and higher in amplitude than the simulated CRN \footnote{Interestingly, we recover similar spectrum for GWB in EPTA DR2new data set \cite{EPTA:2023sfo}.}.

We analysed the same data assuming a CRN+WN model only and overplot the CRN posterior in purple in Fig.~\ref{fig:cgw_crn_corner} which overlaps closely with the orange contours. Therefore, the low-likelihood mode in the full posterior corresponds to the partial absorption of the eCGW by the CRN. We also evaluate the free spectrum of our simulated data, presented in Fig.~\ref{fig:cgw_crn_corner}. 
\begin{figure}[h]
        \includegraphics[width=0.5\textwidth]{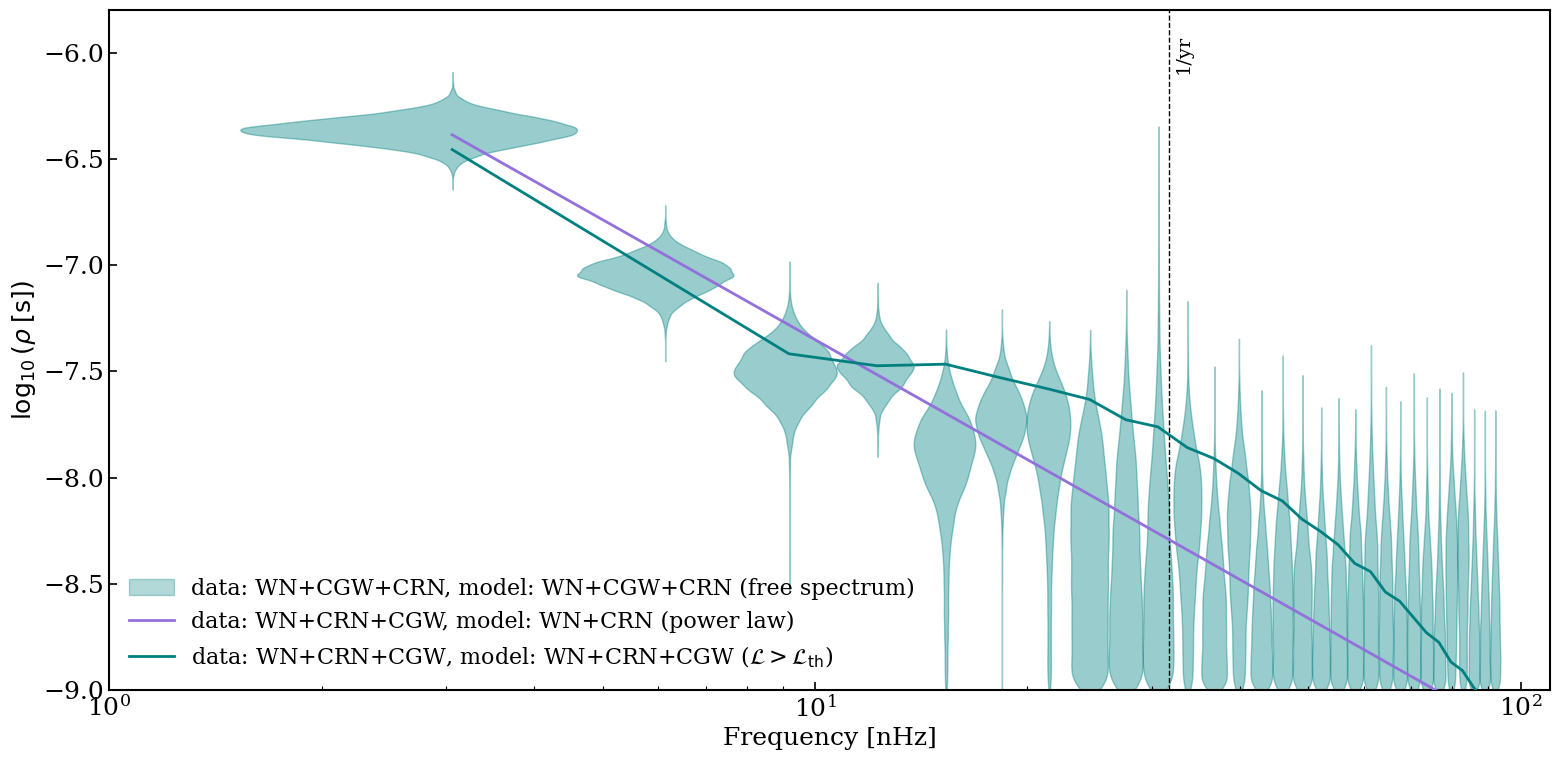}
        \caption{Teal violin plots represent the free spectrum fit on the dataset containing WN+CRN+eCGW at high frequency, the teal line is the sum of the PSD of the CGW and the CRN obtained when the two signals are properly disentangled. The purple line is the CRN PSD obtained from WN+CRN on the same dataset}
        \label{fig:cgw_crn_spectrum}
\end{figure}
We overplot it with the spectrum corresponding to the injection (CRN+eCGW+WN) in teal, and with the purple line corresponding to the CRN only fit to the same data (the colors match posteriors in Fig.~\ref{fig:cgw_crn_corner}).

These results are consistent with the low Bayes factor of $\mathcal{B}^{\mathrm{CRN+CGW}}_{\mathrm{CRN}} = 0.7$ evaluated using the product space approach (see \cite{Korsakova:2024sut}  and references therein). This indicates no significant preference for the eCGW+CRN model over the CRN-only hypothesis.
The MCMC loses track of the signal and explores regions of parameter space consistent with noise only, suggesting that the noise model provides a comparable  (actually slightly worse) description of the data. This is likely a consequence of the large prior volume associated with the eCGW parameters — 63 parameters for the CGW+CRN model compared to the 2 parameters of the CRN-only model — which penalizes the evidence integral even in the presence of a strong likelihood peak of $\Delta \ln\mathcal{L} = 16.5$. Hence, the CRN-only model cannot be ruled out because of its simplicity (only two parameters) and its similarity to the broadband eCGW spectrum.

\begin{figure}[h!] 
    \includegraphics[width=0.5\textwidth]{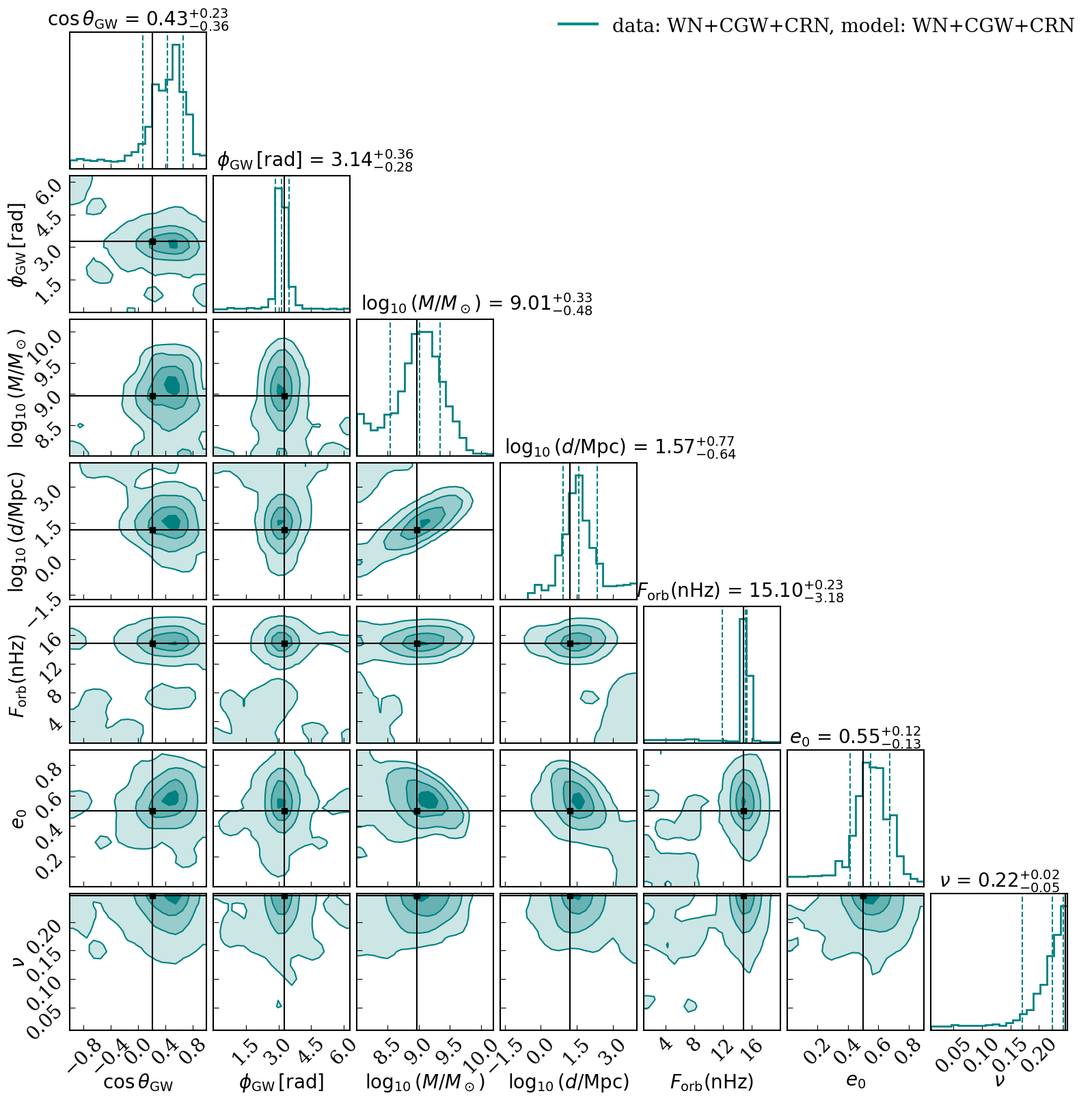} 
    \caption{Corner plot for the high frequency source. Data and model consist of WN+CRN+eCGW.}
    \label{fig:corner_CRN_CGW_highF}
\end{figure}

\vspace{1.5cm}
We now consider an eCGW signal from a binary with a higher orbital frequency ($F_{\mathrm{orb}} = 15 nHz$). The Earth terms fall completely on the WN-dominated part of the spectrum, while the pulsar terms still have some overlap with the CRN. The posteriors are presented in Fig.~\ref{fig:corner_CRN_CGW_highF}. 

\begin{figure}[t]
        \centering
        \includegraphics[width=0.5\textwidth]{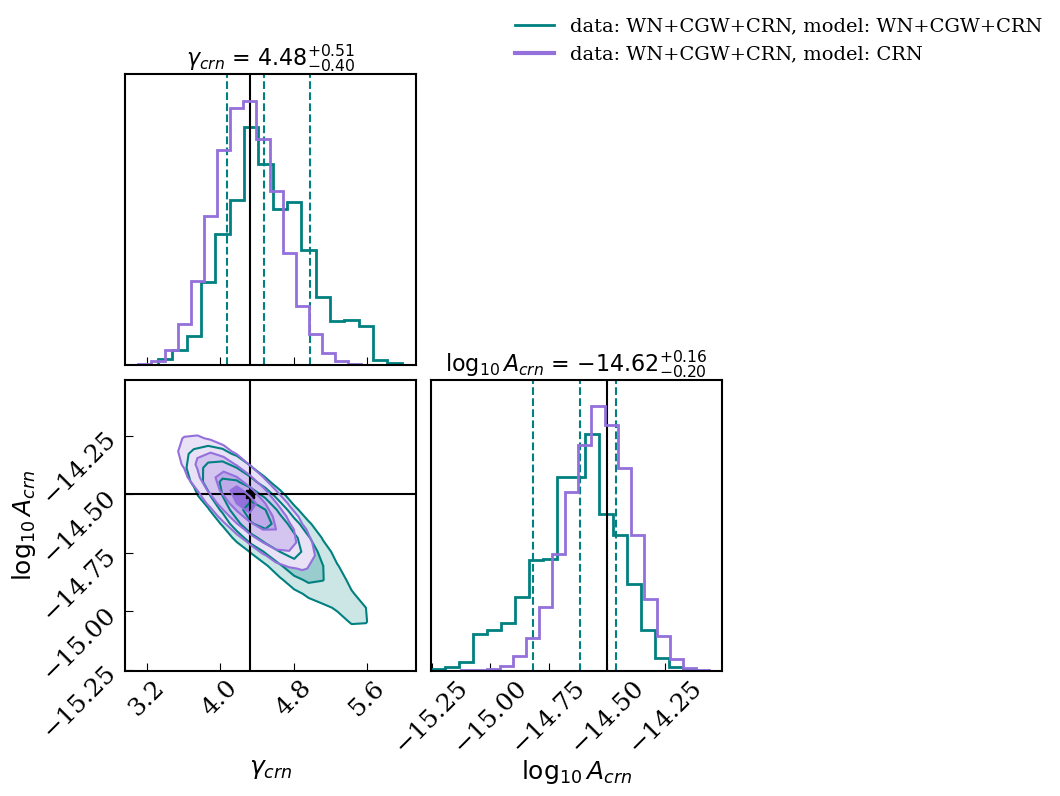}
        \caption{Posterior distribution in teal resulting from fitting dataset with WN+CRN+eCGW at high frequency with model WN+CRN+eCGW compared with the purple posterior obtained from fitting the same dataset with the WN+CRN model.}
        \label{fig:corner_crn_highF}
\end{figure}

\begin{figure}[t]
    \centering
    \includegraphics[width=0.5\textwidth]{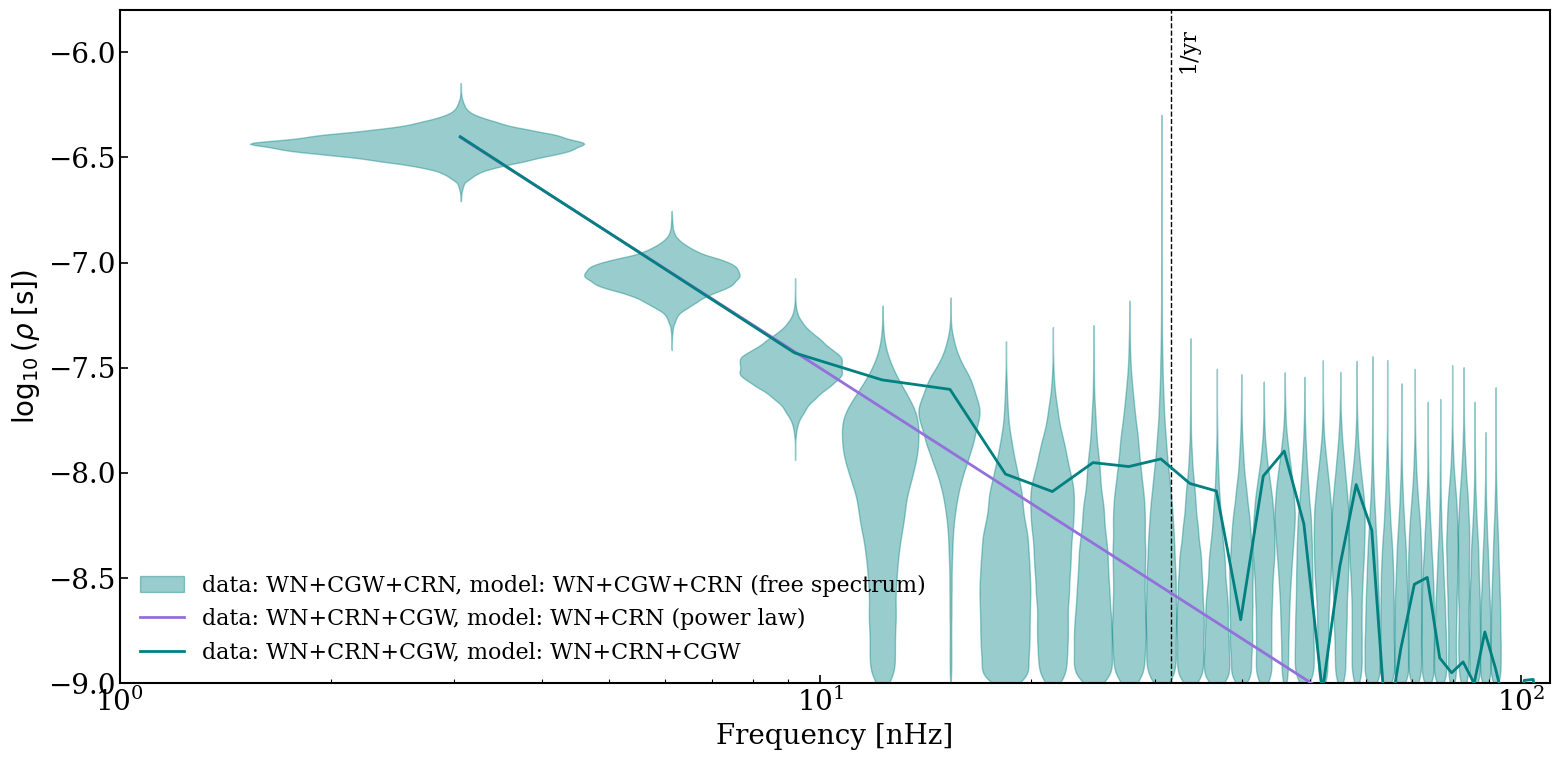}
    \caption{Teal violin plots represent the free spectrum fit on the dataset containing WN+CRN+eCGW at high frequency, the teal line is the sum of the PSD of the CGW and the CRN obtained when the two signals are properly disentangled. The purple line is the CRN PSD obtained from WN+CRN on the same dataset}
    \label{fig:spectrum_highF}
\end{figure}

The absorption of the high-frequency eCGW by the CRN is much less pronounced, since the two processes overlap only in a subset of the pulsar-term harmonics in the frequency domain.  The intrinsic eCGW parameters, namely the orbital frequency $F_{\mathrm{orb}} = 15.10 nHz$, the eccentricity $e_0 = 0.55$, and the symmetric mass ratio $\nu = 0.22$, are well constrained. Nevertheless, we still observe a shallow tail associated with the correlation between the eCGW signal and the CRN. 
The recovered CRN posterior is shown in Fig.\ref{fig:corner_crn_highF}.
As in the low-frequency case, the teal contours correspond to the full CRN+eCGW inference, while the purple contours are obtained by analysing the same simulated data with the CRN-only model. The free spectrum of the simulated data is shown in Fig.\ref{fig:spectrum_highF}. For comparison, we overplot the injected spectrum (teal) and the best-fitting CRN-only spectrum (purple), using the same colour scheme as in the left panel. We observe that the recovered CRN spectrum overlaps with the injected. The evaluated Bayes factor (noise only vs noise  plus eCGW), which is of order unity, does not yet provide strong evidence in favour of the CGW+CRN model over the CRN-only hypothesis. This is primarily due to the high dimensionality of the CGW+CRN parameter space, which penalizes the Bayesian evidence despite the improved likelihood.

The evidence penalty associated with the "curse of dimensionality" can be mitigated when the prior volume is reduced, for example in targeted searches where some prior ranges are constrained by external information.
It is therefore possible to obtain informative posteriors without a large Bayes factor. As an alternative diagnostic for an "emerging eCGW" signal, we evaluate the Hellinger \cite{Bhattacharyya:1943} and Jensen-Shannon \cite{Lin:1991zzm} marginalised per-parameter distances between the posteriors $p(\Theta_i)$ and the priors $\pi(\Theta_i)$ for the full WN+CRN+eCGW runs at low and high frequency. We define these two metrics in the Appendix~\ref{sec:distances}.

\begin{figure*}[t] 
    \includegraphics[width=\textwidth]{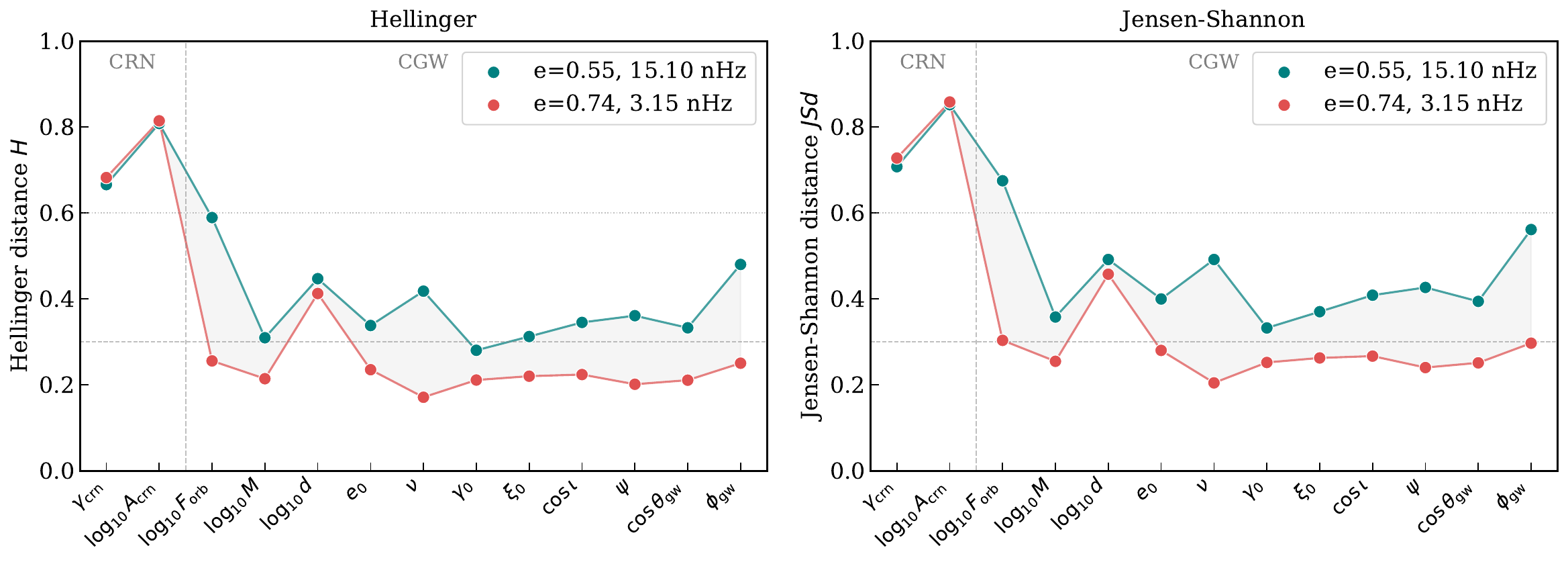} 
    \caption{Hellinger distance (left) and Jensen-Shannon divergence (right) for two 
    eccentric CGW signals: $e_0=0.5$, $F_{\mathrm{orb}}=15\,\mathrm{nHz}$ (teal) 
    and $e_0=0.75$, $F_{\mathrm{orb}}=3\,\mathrm{nHz}$ (red).}
    \label{fig:hellinger}
\end{figure*}

The per-parameter metrics for the eCGW parameters of both binaries considered in this study are presented in Fig.~\ref{fig:hellinger}. The large prior--posterior Hellinger and Jensen-Shannon distances show that the data are informative about the CGW parameters, particularly for the high-frequency eCGW. 
The Hellinger and Jensen-Shannon diagnostics show similar qualitative trends, but probe the prior--posterior difference in complementary ways: the Hellinger distance measures geometric overlap, while the Jensen-Shannon divergence quantifies information gain.  
This highlights the limitation of Bayes factors in high-dimensional parameter-estimation problems and motivates parameter-level diagnostics as complementary measures of signal informativeness.

\section{Discussion/Conclusion}\label{sec:discussion}

In this work, we use Bayesian approach to infer parameters of eccentric 
supermassive black hole binaries in the simulated PTA data. We believe, this work is the most
complete investigation of eCGW signal within PTA data. Here we briefly summarize the main conclusions of this paper.

\paragraph{Importance of post-Newtonian corrections and pulsar term:}

Our results confirm the conjecture put forward in~\cite{Manzini:2025gjx}: including higher-order post-Newtonian corrections in the binary evolution is essential for unbiased parameter recovery when the source is at high orbital frequency. In this regime, the binary evolves significantly over the Earth--pulsar light-travel time $\tau_\alpha$, and truncating the dynamics at leading order make posteriors broader and introduces biases. These biases propagate into the estimates of the individual companion masses. 

The pulsar term plays an equally critical role at high orbital frequencies. The Earth-term-only model is unable to reproduce the full harmonic structure of the eCGW signal and leads to unconstrained posteriors for most source parameters. This contrasts with the low-frequency, nearly circular case ($F_\mathrm{orb} = 5,\mathrm{nHz}$, $e_0 = 0.01$), where the binary evolves only mildly over $\tau_\alpha$. In that case, the Earth-term-only model still recovers the frequency, eccentricity, and sky location with acceptable accuracy, although the total mass, distance, and mass ratio remain unconstrained, as expected from the absence of frequency-evolution information provided by the pulsar term.

The ability to measure the individual companion masses $m_1$ and $m_2$, rather than only the chirp mass, is a distinctive feature of the eccentric model. The symmetric mass ratio $\nu$ is constrained through the combination of periapsis precession and the harmonic structure of the waveform, both of which depend sensitively on $\nu$. This measurement becomes possible only when both the pulsar term and high-PN corrections are included, since the individual mass estimates rely on the total mass and symmetric mass ratio inferred from the eccentricity and frequency evolution over $\tau_\alpha$.

Reducing the pulsar distance uncertainty from $20\%$ to $5\%$ primarily improves the sky localisation of the source, while leaving the other binary parameters largely unaffected. This suggests that, at current levels of pulsar distance knowledge, the dominant limitation on sky localisation is the prior on $d_\alpha$ rather than the signal itself. Future astrometric measurements, such as those expected from VLBI campaigns, could therefore meaningfully tighten the source-position constraints.

\vspace{0.5cm}
\paragraph{Correlation between eCGW and CRN:}

The second set of results addresses the interplay between the eCGW signal and the common red noise. At low orbital frequencies ($F_\mathrm{orb} = 3,15\mathrm{nHz}$), the eCGW harmonics overlap substantially with the CRN-dominated frequency band. The resulting degeneracy between the two processes appears as a bimodal likelihood distribution.
In the high-likelihood mode, the eCGW signal is identified and the CRN is recovered consistently with the injection. In the second mode, part of the eCGW power is absorbed by the CRN, which leads to shifts toward a higher amplitude and a shallower spectral index in order to account for the excess power at 
intermediate frequencies. This behaviour has a clear physical interpretation: a highly eccentric binary distributes its GW power across many harmonics producing a broad-band spectrum that can resemble a smooth red process at the frequency resolution of current PTA datasets.

For this injection, the Bayes factor is $\mathcal{B}_\mathrm{CRN}^\mathrm{CRN+CGW} \approx 0.7$, indicating no significant preference for the eCGW signal model over the CRN-only hypothesis. This result is driven by the large prior volume of the eCGW+CRN model. Although the likelihood exhibits a clear peak ($\Delta\ln\mathcal{L} = 16.5$), the 63-parameter eCGW+CRN model is strongly penalized by the evidence integral relative to the 2-parameter CRN model. Thus, the data are informative about parts of the signal parameter space, but not sufficiently so to overcome the Occam penalty associated with the full eCGW model. This illustrates an important limitation of Bayes factors in high-dimensional, low-SNR searches for individual PTA sources.

At higher orbital frequencies ($F_\mathrm{orb} = 15,10\mathrm{nHz}$), the situation improves considerably. Many of eCGW harmonics lie well above the CRN-dominated band, reducing the spectral overlap and the degeneracy between the two processes. The misspecified WN+CRN fit to the WN+CGW+CRN dataset shows only a marginal bias in amplitude and spectral index, and the free-spectrum reconstruction closely tracks the true noise power law across all frequencies. The eCGW posteriors from the joint WN+CGW+CRN run recover the injected orbital frequency, eccentricity, and symmetric mass ratio with good accuracy, while the sky position, chirp mass, and luminosity distance remain broad but still informative. The Bayes factor, however, is of order unity in this case as well, again due to the dimensionality penalty.

The prior--posterior Hellinger distances and Jensen-Shanon divergence shown in Fig.~\ref{fig:hellinger} provide a complementary diagnostic. They show that the data can be informative about the eCGW parameters, especially for the high-frequency source, even when the global Bayes factor does not favour the more complex model. These results motivate the use of parameter-level diagnostics together with global model-selection metrics when assessing candidates for individual sources in PTA data.

\vspace{0.5cm}
\paragraph{Limitations.}

Several simplifications adopted in this work should be kept in mind. First, we model the SGWB as a common uncorrelated red-noise process and neglect the Hellings--Downs spatial correlations. This approximation is not expected to change our main conclusions about eCGW parameter recovery. However, a fully spatially correlated background could, in principle, produce stronger degeneracies with highly eccentric, low-frequency eCGWs than the CRN model used here.

Second, we did not include individual pulsar red noise or chromatic noise components. In an ideal dataset with sufficiently broad radio-frequency coverage, chromatic noise should decouple from achromatic red noise. This was not fully the case for EPTA DR2, but future data releases including LOFAR and NenuFAR observations should help address this limitation. Individual pulsar red-noise components are expected to have an effect qualitatively similar to that of the CRN.

Third, although the simulated dataset follows realistic EPTA DR2new configurations, we fix the CRN and WN parameters. In reality, these parameters are drawn from the broad posterior distributions reported in EPTA studies \cite{EPTA:2023sfo, EPTA:2023fyk}. 

Finally, on the signal-modelling side, the numerical integration of the binary evolution equations over $\tau_\alpha$ remains the main computational bottleneck in the evaluation of the pulsar term. Faster approximations, for example, based on neural-network-generated waveforms, would significantly improve the efficiency and extend the reach of the current pipeline.

\acknowledgements
This work was supported by the French National Research Agency (grant ANR-21-CE31-0026, project MBH\_waves).
We also acknowledge financial support from the Centre national d'études spatiales (CNES), France (ROR: https://ror.org/04h1h0y33), within the framework of the LISA space mission. We thank Irene Ferranti and Riccardo Jurgen Truant for their help during the preparation of this work.



\appendix
\section{Appendix: Jensen-Shannon and Hellinger distances}\label{sec:distances}
To quantify the level of agreement between posterior distributions obtained under different waveform models, we employ two complementary statistical distance measures: the Jensen–Shannon (JS) distance and the Hellinger distance. Both are symmetric, bounded metrics that capture the full shape of a distribution, including differences in width, skewness, and multimodality that may not be evident from percentile or credible-interval comparisons alone.
The Hellinger distance is expressed as

\begin{equation}\label{eq:hellinger_d}
H(p(\Theta_i), \pi(\Theta_i)) = \frac{1}{\sqrt{2}}\sqrt{\sum_{j=1}^{N_{\mathrm{samples}}}\left(\sqrt{p(\Theta_{i,j})} -\sqrt{\pi(\Theta_{i,j})}\right)}
\end{equation}
while the Jensen-Shannon divergence is given by:
\begin{equation}\label{eq:jensen-shannon}
\mathrm{JSD}(p(\Theta_i) \| \pi(\Theta_i)) = \frac{1}{2}\left( \mathrm{KL}(p(\Theta_i) \| m(\Theta_i)) +  \mathrm{KL}(\pi(\Theta_i) \| m(\Theta_i))\right)
\end{equation}
where $m(\Theta_i)$ is the average between per-parameter posterior and prior:
\begin{equation}\label{eq:jensen-shannon-mixture}
m(\Theta_i) = \frac{1}{2}(p(\Theta_i) + \pi(\Theta_i))
\end{equation}
and KL is the Kullback-Leibler divergence
\begin{equation}\label{eq:kl-divergence}
D_{\mathrm{KL}}(p \| q) = \sum_{j=1}^{N_{\mathrm{samples}}} p(\Theta_{i,j}) \log\frac{p(\Theta_{i,j})}{q(\Theta_{i,j})}.
\end{equation}
The JS distance is defined as a root square of JS divergence; it behaves like a proper distance, including satisfying the triangle inequality.

\section{Appendix: Prior Distributions}
Table \ref{tab:priors} summarizes the prior distributions adopted for all parameters in the parameter estimation analysis, including pulsar distances, true anomalies, common red noise parameters, and continuous-wave source parameters.
\begin{table*}[h]
\centering
\small
\begin{tabular}{llll}
\toprule
\textbf{Parameter} & \textbf{Distribution} & \textbf{Property 1} & \textbf{Property 2} \\
\midrule
\multicolumn{4}{l}{\textit{Pulsar Distance Parameters}} \\
\midrule
J0030+0451 distance & Normal & $\mu = 0.28$ & $\sigma = 0.10$ \\
J0613$-$0200 distance & Normal & $\mu = 0.90$ & $\sigma = 0.40$ \\
J1012+5307 distance & Normal & $\mu = 0.70$ & $\sigma = 0.20$ \\
J1022+1001 distance & Normal & $\mu = 0.52$ & $\sigma = 0.09$ \\
J1024$-$0719 distance & Normal & $\mu = 0.49$ & $\sigma = 0.12$ \\
J1455$-$3330 distance & Normal & $\mu = 0.74$ & $\sigma = 0.15$ \\
J1600$-$3053 distance & Normal & $\mu = 2.40$ & $\sigma = 0.90$ \\
J1640+2224 distance & Normal & $\mu = 1.19$ & $\sigma = 0.238$ \\
J1713+0747 distance & Normal & $\mu = 1.05$ & $\sigma = 0.06$ \\
J1730$-$2304 distance & Normal & $\mu = 0.51$ & $\sigma = 0.10$ \\
J1744$-$1134 distance & Normal & $\mu = 0.42$ & $\sigma = 0.02$ \\
J1857+0943 distance & Normal & $\mu = 0.90$ & $\sigma = 0.20$ \\
J1909$-$3744 distance & Normal & $\mu = 1.26$ & $\sigma = 0.03$ \\
J1910+1256 distance & Normal & $\mu = 1.95$ & $\sigma = 0.39$ \\
J1918$-$0642 distance & Normal & $\mu = 1.40$ & $\sigma = 0.28$ \\
J2124$-$3358 distance & Normal & $\mu = 0.30$ & $\sigma = 0.07$ \\
J0751+1807, J0900$-$3144, J1738+0333, etc. & Normal & $\mu = 1.00$ & $\sigma = 0.20$ \\
\midrule
\multicolumn{4}{l}{\textit{True Anomalies at Pulsars}} \\
\midrule
$\xi_{0,p}$ & Uniform & $\text{min} = -\pi$ & $\text{max} = \pi$ \\
\midrule
\multicolumn{4}{l}{\textit{Common Red Noise Parameters}} \\
\midrule
$\gamma_\mathrm{CRN}$ & Uniform & $\text{min} = 0$ & $\text{max} = 7$ \\
$\log_{10} A_\mathrm{CRN}$ & Uniform & $\text{min} = -18$ & $\text{max} = -10$ \\
\midrule
\multicolumn{4}{l}{\textit{Continuous Wave (GW Source) Parameters}} \\
\midrule
$\cos\theta_\mathrm{GW}$ & Uniform & $\text{min} = -1$ & $\text{max} = 1$ \\
$\cos\iota$ & Uniform & $\text{min} = -1$ & $\text{max} = 1$ \\
$e_0$ & Uniform & $\text{min} = 0.001$ & $\text{max} = 0.9$ \\
$\gamma_0$ & Uniform & $\text{min} = 0$ & $\text{max} = \pi$ \\
$\phi_\mathrm{GW}$ & Uniform & $\text{min} = 0$ & $\text{max} = 2\pi$ \\
$\log_{10} F_\mathrm{orb}$ & Uniform & $\text{min} = -9.0$ & $\text{max} = -7.7$ \\
$\log_{10} M$ & Uniform & $\text{min} = 8.0$ & $\text{max} = 10.2$ \\
$\log_{10} d$ & Uniform & $\text{min} = -2$ & $\text{max} = 4$ \\
$\nu$ & Uniform & $\text{min} = 0.01$ & $\text{max} = 0.25$ \\
$\psi$ & Uniform & $\text{min} = 0$ & $\text{max} = \pi$ \\
$\xi_0$ & Uniform & $\text{min} = -\pi$ & $\text{max} = \pi$ \\
\bottomrule
\end{tabular}
\caption{Prior distributions for all parameters used in parameter estimation.}
\label{tab:priors}
\end{table*}

\clearpage
\bibliographystyle{apsrev4-2}
\bibliography{clean_inspire_style_references}

\end{document}